\pdfoutput=1 
\documentclass[12pt,twoside]{article}
\def\neave{<n_e>\;}
\def\sun{\hbox{$\odot$}}
\def\h0units{\mathrm{km\,s^{-1}\,Mpc^{-1}}}
\def\neunits{\mathrm{particles \, cm^{-3}}}

\def\sun{\hbox{$\odot$}}
\date{}
\usepackage {graphicx} 
\def\aj{AJ  }%% The Astronomical Journal
\def\apj{ApJ\,  }%% Astrophysical Journal
\def\apss{Astrophysics and Space Science  }%Astrophysics and Space Science
\def\mnras{MNRAS\,  }%% Monthly Notices of the RAS
\def\na {New Astronomy\,  }%% New Astronomy
\def\pra{Phys. Rev. A   }% % Physical Review A
\def\prd{Phys. Rev. D   }% % Physical Review D
\def\pre{Phys. Rev. E   }% % Physical Review E
\begin{document}
\centerline{\bf Adv. Studies Theor. Phys, Vol. x, 200x, no. xx,
xxx - xxx}

\centerline{}

\centerline{}

\centerline 
{
Anisotropy in the Hubble constant
as modeled by density gradients
}
%% Place for inserting article cathegory: Research Article, Rapid Communication, Communication or Review Article
\author
{
L. Zaninetti   \\
Dipartimento di Fisica \\
Via Pietro Giuria 1    \\
10125, Turin, Italy    \\
}
\centerline{}

\centerline{\bf {L. Zaninetti}}

\centerline{}

\centerline{Dipartimento  di Fisica,}

\centerline{Universit\`a degli Studi di Torino,}

\centerline{via P. Giuria 1,  10125 Torino, Italy}

\begin{abstract}

The all-sky maps of the observed variation of  the
Hubble constant  can be reproduced from 
a theoretical point of view
by introducing an intergalactic plasma with
a variable number density of  electrons.
The  observed averaged value and variance 
of the Hubble constant
are reproduced by adopting
a  rim model,
an auto-gravitating  model,
and  a Voronoi diagrams model
as the backbone  for an auto-gravitating medium.
We also analyze   an
astronomer's model   based on the
3D spatial  distribution of
galaxies as given by
the 2MASS Redshift Survey
and  an auto-gravitating
Lane--Emden ($n=5$) profile of the electrons.
The simulation which involves
the Voronoi diagrams  is done in a cubic
box with sides of 100 Mpc.
The simulation
which involves  the 2MASS
covers the range of redshift smaller
than 0.05. 
\end{abstract}

\section{Introduction}

Hubble's constant \cite{Hubble1929}
is characterized at the moment of writing
by  a large uncertainty.
A recent  evaluation,
see \cite{Freedman2012},
quotes
\begin{equation}
H_0 =(74.3 \pm 2.1 ) \h0units
\quad ,
\end {equation}
which means a relative standard uncertainty of
28264 parts per
million (in the following ppm).
As a comparison,
the value of the Newtonian gravitational constant, denoted by $G$,
is
\begin{equation}
G= (6.67384\,10^{-11} \pm 0.0008) m^3 kg^{-1} s^{-2}
\quad ,
\end{equation}
which means  an uncertainty of 120 ppm, see  \cite{CODATA2012}.
Hubble's constant  is actually  much used
in the series of models starting with
\cite{Friedmann1922,Friedmann1924} as well
as in the modern theories on the accelerating universe,
see  \cite{Riess1998,Perlmutter1999}.
In this standard  cosmology the Hubble parameter is
defined as
\begin{equation}
H =\frac{\dot{a}}{a}
\quad ,
\end{equation}
where a(t) is the scale factor.
The currently  observable value of the Hubble constant
is H$_0$
and in the standard cosmology  is
independent  of the chosen line of sight
of the observer situated on the Earth.
A different  explanation for 
Hubble's constant  lies in the momentum
lost by the light  during the travel in the intergalactic
medium (IGM).
The first   formula for the change
of frequency of the light in a gravitational framework
was due  to  \cite{Zwicky1929}:
\begin{equation}
\frac {\Delta \nu} {\nu}  =
\frac{1.4  \pi G \rho D L}{c^2}
\quad .
\end{equation}
Here,
$\nu$  is the considered frequency,
$G$    is the Newtonian gravitational constant,
$\rho$ is the density in g/cm$^3$,
$D$    is the distance after which  the perturbing effect
       begins to fade out,
$L$    is the considered distance, and
$c$    is the speed of light.
The study  of the physical mechanisms which produce
the redshift of light  was  dormant for 70 years
but currently is explained  by different  physical models.
We list some of the processes which produce
the observed redshift of galaxies:
\begin{itemize}
\item  a plasma  physics effect, see
                 \cite{Brynjolfsson2004,Brynjolfsson2009},
\item a photo-absorption process, see
                 \cite{Ashmore2006},
\item an interaction of photons with curved space time,
      see  \cite{Crawford2006,Crawford2011},
\item an  interaction of photons
          with intergalactic free electrons,
          see \cite{Mamas2010},

\item an interaction between single photons
      traveling across the micro-quanta flux,
      see  \cite{Michelini2013}.
\end{itemize}
The detailed analysis of these and other
physical mechanisms which produce
the observed redshift can be
found in \cite{Marmet2009}.

The explanation of the anisotropies in
Hubble's constant
started  with \cite{Karoji1975}
analyzing  the Markarian galaxies.
The deviations from the Hubble flow
for  the brightest galaxies were analyzed in
\cite{Guthrie1976}.
A self similar cosmology was applied to explain the
anisotropies in Hubble's constant, see
\cite{Fennelly1977}.
The radial and angular variance in  the Hubble flow
are derived  on scales in which the flow is in
the nonlinear regime, see \cite{Wiltshire2013}.
In this
paper, Section
\ref{sectionpreliminaries}
reviews the existing
data on the anisotropy  of the Hubble constant
and two plasma effects within the
IGM
which produce
the observed redshift.
Section \ref{standard} reports
the standard approach to the accelerating universe.
Section \ref{nonhomo} reviews the current
status of the astronomical  observations which point towards
a non-homogeneous  local universe.
Section \ref{sectiontwo} introduces two
simple models based on the
average value of the cosmic voids
which produce  an anisotropy
in the  Hubble constant.
Section \ref{sectiontheoretical}
introduces a
Voronoi diagrams network  for the 3D spatial
distribution of galaxies  and
a logistic distribution
for the number density of electrons,
which  leads to an anisotropy
in the Hubble constant.
Section \ref{sectionastronomers} inserts
the  3D spatial distribution of galaxies
as given   by a  recent  catalog
a number density profile   of  electrons
of  the Emden type:  as a consequence,
contour maps of the Hubble constant are generated.

\section{Standard Cosmology}
\label{standard}

\begin{figure*}
\begin{center}
\includegraphics[width=10cm]{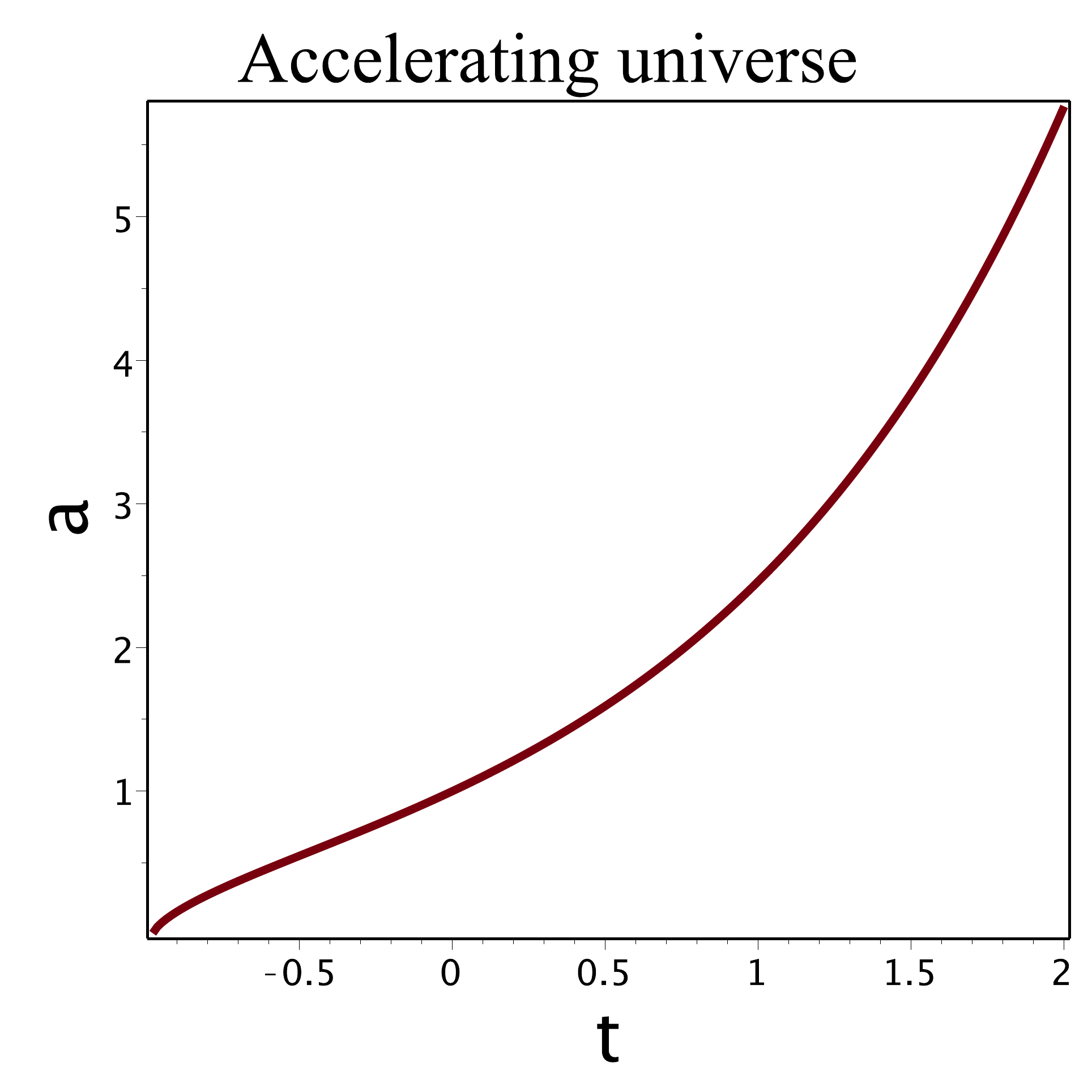}
\end {center}
\caption
{
Evolution of the scale factor from the singularity at $t=-1$,
to $t=2$.
}
\label{acceleration}
    \end{figure*}

The field equations in general relativity (GR), after
\cite{Einstein1916}, are
\begin{equation}
\label{eqngr}
R_{\mu\nu} - \frac{g_{\mu\nu}R}{2} =
-\kappa \,T_{\mu\nu}
\end{equation}
with
\begin{equation}
\kappa  = \frac{8 \pi G}{c^4}
\quad ,
\end{equation}
where
$R_{\mu\nu}$  is the Ricci tensor of the first kind,
$g_{\mu\nu}$  is the metric tensor,
$T_{\mu\nu}$  is the energy--momentum tensor,
R             is the Ricci scalar,
G             is the gravitational constant,
and  c        is the speed of light.
The introduction of the cosmological  constant $\Lambda$,
see \cite{Einstein1917,Einstein1922},
transforms  the previous  equations  into
\begin{equation}
\label{eqncosmo}
R_{\mu\nu} - \frac{g_{\mu\nu}R}{2} -\Lambda g_{\mu\nu} =
-\kappa \,T_{\mu\nu} \quad .
\end{equation}
On inserting  the  Friedmann--Robertson--Walker (FRW) metric
and the perfect fluid tensor into the Einstein equations, we obtain
the  Friedmann equation
\begin{equation}
\left({\dot a(t)\over a(t)}\right)^2 = {8\pi G\over3}\,\rho +
{\Lambda c^2\over3} - {\kappa c^2\over a(t)^2 }
\quad ,
\end{equation}
where $\rho$ is the  density,
$\kappa$ takes the values -1,0,1 for negative, zero, or positive spatial curvature,
and $a(t)$ is the scale factor,
see \cite{Friedmann1922,Friedmann1924}.
The previous equation  has been deduced
calculating the  volume, $V$, of the 3D hypersphere
\begin{equation}
V=2 \, \pi^2 \, a(t)^3
\end{equation}
and the total mass, M,
\begin{equation}
M = \rho \, V
\quad .
\end{equation}
This means that the local universe is assumed to 
be homogeneous, $\rho$ is constant, and isotropic, 
the integration for the volume  over the 
three angles is standard.
The study of 42 supernovae of type Ia  in \cite{Perlmutter1999}
suggested an expanding universe with acceleration.
The acceleration can be explained by introducing the dark matter
parameter $\Omega_{\Lambda}$, where the index $\Lambda$ indicates its similarity
with the cosmological constant.
The Friedmann equation now  is
\begin{equation}
\frac{\dot{a}^2}{a^2}
+ \frac{H_0^2 (\Omega_0 -1) } {a^2}= H_0^2 (\frac{\Omega_m}{a^3}
+ \frac{\Omega_r}{a^4}
+ \Omega_{\Lambda}  )
\quad ,
\label{eqnacceleration}
\end{equation}
where the actual energy density parameter,
$\Omega_0$,   is
given by the sum of matter, radiation, and dark components
\begin{equation}
\Omega_0 =
\Omega_m +
\Omega_r +
\Omega_{\Lambda}
\end{equation}
and
$H_0$ is the current value of the Hubble constant,
see \cite{Wang2006} for more details.
A recent evaluation gives a dark energy density $\Omega_{\Lambda}$ =0.7185,
a matter density $\Omega_m$ =0.235,
$H_0= 69.7 ,\h0units$,
and age of the universe $t_0$ = 13.76\,Gyr.
With these parameters, Figure \ref{acceleration}
shows  the temporal behavior
of the scale factor as given by the solution of
(\ref{eqnacceleration}).
%begin figure acceleration
%end  figure  acceleration

\section{The non-homogeneous universe}

\label{nonhomo}
\begin{figure*}
\begin{center}
\includegraphics[width=10cm]{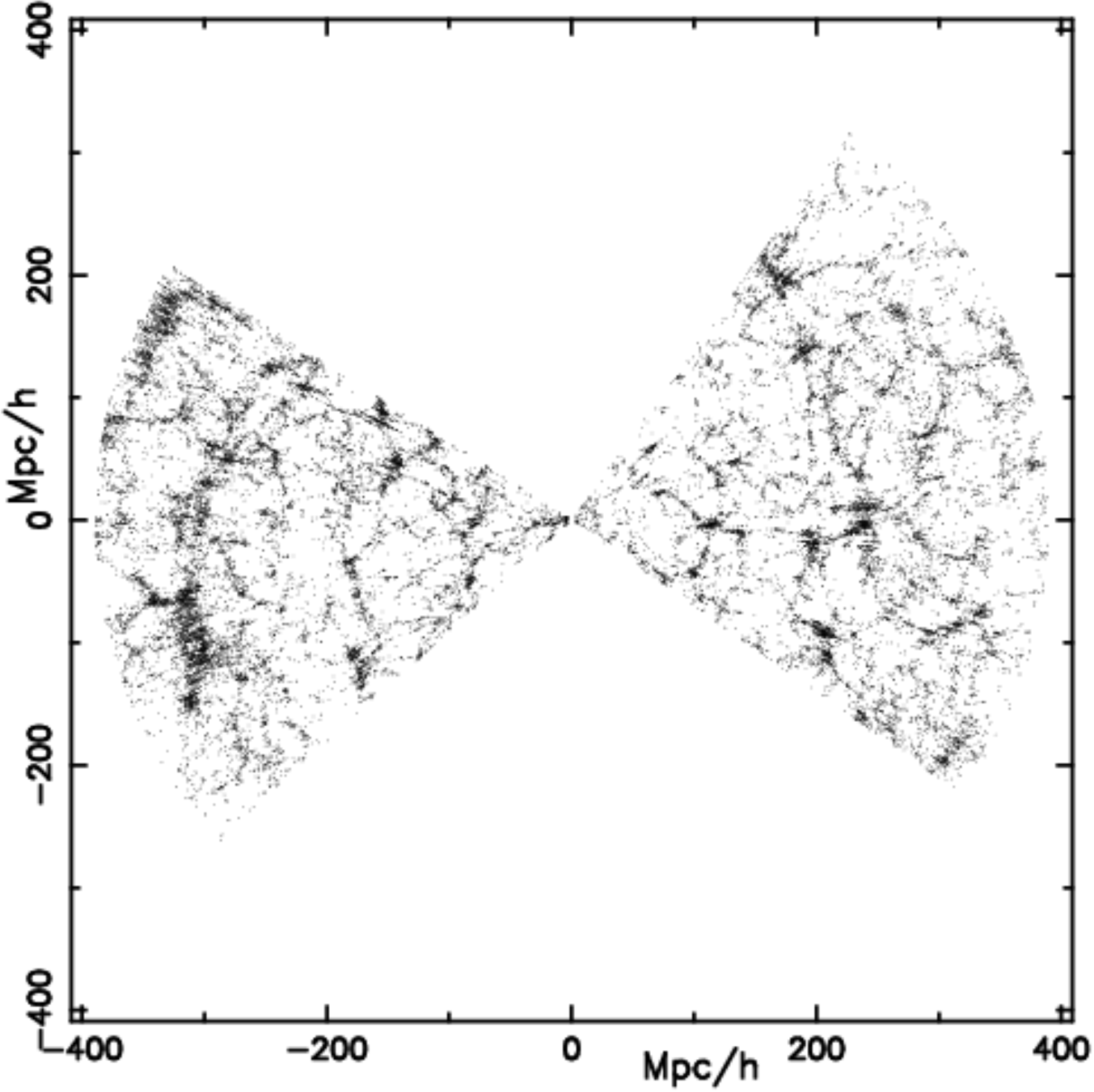}
\end {center}
\caption{Slice  of   $75^{\circ} \times 3^{\circ}$
in the 2dFGRS.
This plot contains  30000  galaxies and belongs to the
 2dFGRS
available at  http://msowww.anu.edu.au/2dFGRS/  \,.
}
          \label{2dfslice}%
    \end{figure*}

\begin{figure}
\begin{center}
\includegraphics[width=6cm]{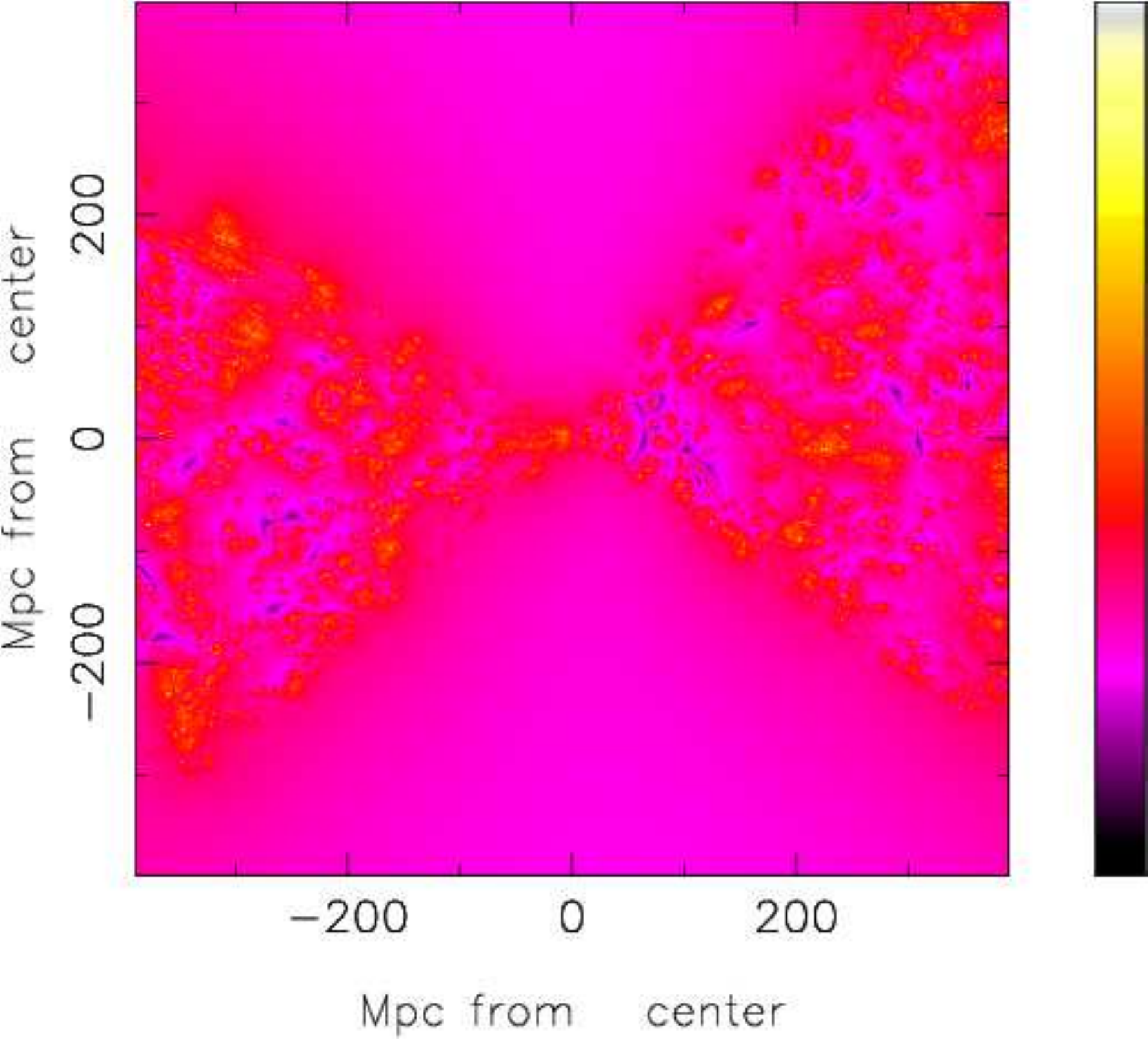}
\end {center}
\caption
{
Map of the decimal  logarithm of the modulus
of the gravitational force for 2dFGRS.
}
          \label{2dfieldmap}%
    \end{figure}

\begin{figure}
\begin{center}
\includegraphics[width=6cm]{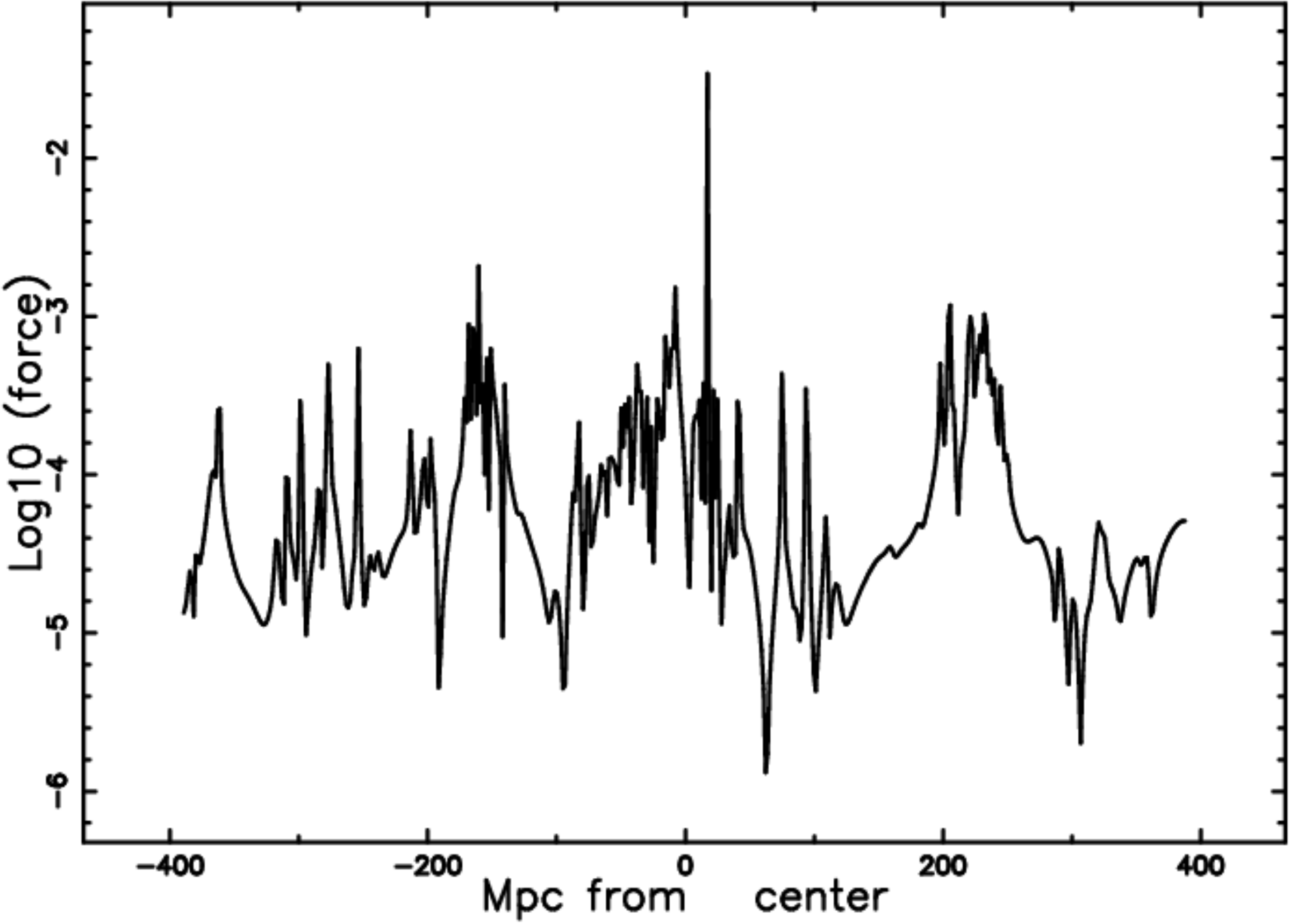}
\end {center}
\caption
{
Cut along  the center for  the decimal  logarithm of the modulus
of the gravitational force for 2dFGRS.
}
          \label{2dfieldcut}%
    \end{figure}

The standard cosmology is based on the concept of a homogeneous
and isotropic universe.
Recent observations  point 
instead towards a cellular structure
of the local universe. 
As an example, we  report
the 2dF Galaxy Redshift Survey (2dFGRS) catalog    
when a slice of
$75^{\circ} \times 3^{\circ}$
is considered,
see Figure \ref{2dfslice}.
%begin figure 2dfslice
%end   figure 2dfslice
The observational fact that the spatial position
is digitized allows us to map
the Newtonian gravitational field.
We firstly express the
Newtonian
gravitational constant  in the following astrophysical units:
length  in Mpc,  mass in M$_{\mbox{gal}}$ which is
$10^{11} M_{\sun}$  and  yr$_8$
which are $10^8$ yr:
\begin{equation}
G=4.49975  \,10^{-6}
\frac { \mbox{Mpc}^3} {M_{\mbox{gal}}\mbox{yr}_8^2}
\quad .
\end{equation}
To each galaxy we associate a mass of $10^{12} M_{\sun}= 10 M_{\mbox{gal}} $
and we evaluate the gravitational force,
expressed in $\frac { \mbox{Mpc}  M_{\mbox{gal}}}{\mbox{yr}_8^2}$, on a unitarian mass, 1$M_{\mbox{gal}}$.
For practical  purposes in each point of a 2D grid
made by 500$\times$500 points we compute the gravitational
force, $\overrightarrow{F}$ by
\begin{equation}
\overrightarrow{F} = -\sum_{i=1}^N  G \frac{M_i}{r_i^3} \overrightarrow{r}
\quad ,
\end{equation}
where the index $i$ runs from 1 to the 
number of considered galaxies, $N$,
$M_i$ is the mass of the considered galaxy,  $10 M_{\mbox{gal}}$,
$\overrightarrow{r}$ is the distance between the grid's point
and the considered galaxy,
and $G$ is the Newtonian constant of gravitation  expressed
in $\frac { \mbox{Mpc}^3} {M_{\mbox{gal}}\mbox{yr}_8^2}$.
Figure \ref{2dfieldmap} reports a 2D map
of the modulus of such force and Figure \ref{2dfieldcut},
a cut in the middle of such a map.
%begin figure 2dfieldmap
% end figure 2dfieldmap

%begin figure 2dfieldcut
% end figure 2dfieldcut
A visual inspection of the previous two figures 
reveals   great
variations in the gravitational  forces and 
a minimum value
at the center of the voids.

%siamoqui
\section {Preliminaries for the Hubble constant}

\label{sectionpreliminaries}

This section reviews the  anisotropy  in the Hubble constant
and two physical mechanisms that
model the Hubble constant.
\subsection{All sky data}
The \textit{HST} Key Project has explored
the value of  $H_0$ in all
directions of the sky and the data
are reported in Tables 1 and 2
of \cite{McClure2007}.
Table~\ref{hubblemanyhst} reports
the average,
the standard deviation,
the weighted mean,
the error of the weighted mean,
the minimum, $H_{0,min}$  and
the maximum  $H_{0,max}$  computed as  in
\cite{Leo1994,Zaninetti2010c}.

\begin{table}[ht!]
\caption {
The Hubble constant
of  the \textit{HST} Key Project
}
\label{hubblemanyhst}
\begin{center}
\begin{tabular}{|c|c|c|}
\hline
entity & definition & value   \\
\hline
$n$ & No. of samples    & 76 \\
$\bar {x}$ & average & 76.7            $\h0units$ \\
$\sigma $ & standard~deviation & 10.57 $\h0units$ \\
$H_0,max$ & maximum & 124.4            $\h0units$    \\
$H_0,min$ & minimum & 54.79            $\h0units$ \\
$\mu$ & weighted~mean &  72.09         $\h0units$\\
$\sigma(\mu)$ & error~of~the~weighted~mean & 0.41
                                       $\h0units$ \\
\hline
\end{tabular}
\end{center}
\end{table}

Our subsequent simulations will   be
calibrated based on this table.

\subsection{Hubble's law}

\begin{figure*}
\begin{center}
\includegraphics[width=10cm]{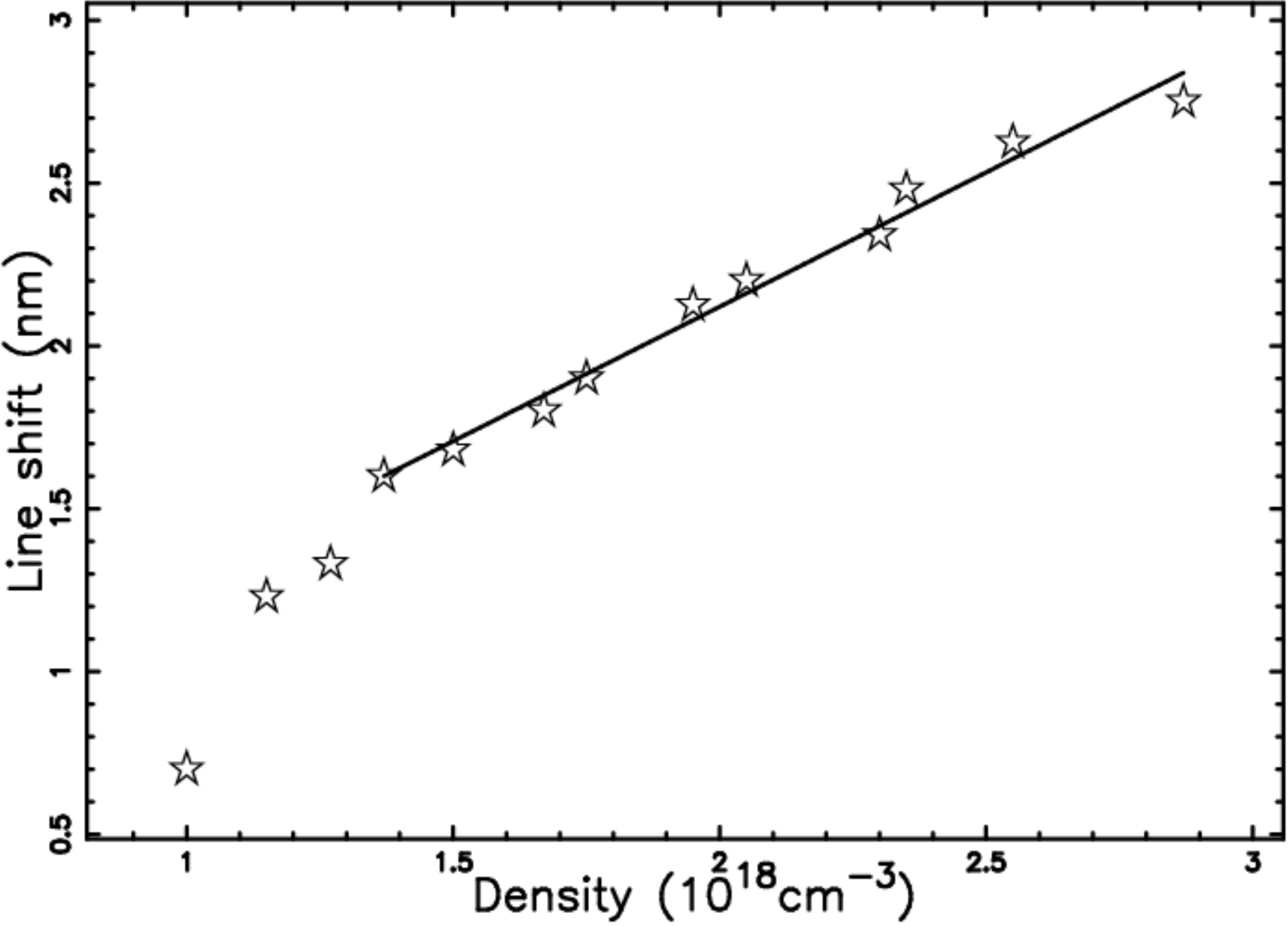}
\end {center}
\caption
{
HgI 435.83 nm line shifts
versus the electron density,
data as extracted  by the author
from   Figure 7 in \cite{Chen2009} (empty stars)
and linear regime                   (full line).
}
\label{lineshift}%label
    \end{figure*}

Starting from \cite{Hubble1929}, the suggested
correlation between
the expansion velocity and distance
in the framework of the Doppler
effect is
\begin {equation}
V= H_0 D = c \, z
\quad ,
\end{equation}
where $H_0$ is the Hubble constant,
$H_0 = 100 h$ $\h0units$,
with $h=1$ when $h$ is not specified,
$D$ is the distance in Mpc,
$c$ is the speed of light and $z$ the redshift.
The Doppler effect produces a linear
relation between distance and redshift.
The analysis of the physical
mechanisms which predict a direct relation
between distance
and redshift started with
\cite{Marmet1988} and a current list of
the various mechanisms can be found in
\cite{Marmet2009}. Here,
we select two mechanisms.
The first mechanism works  in the framework
of a
hot plasma with low density,
such as in the IGM, and
produces a relation of the type
\begin{equation}
D = \frac{{3.0064 \cdot 10^{24} }}{{ \neave
}}\ln \left( {1 + z} \right)~~{\rm{cm }}{\rm{}} \quad ,
\end{equation}
where  the averaged density of electrons,
$\neave$ as expressed in CGS, is
\begin{equation}
\neave  = \frac{{H_0}}{{3.077 \cdot 10^5 }} \frac{\mbox{particles}}{\mbox{cm}^3}
\quad  ,
\end{equation}
see equations (48) and (49) in \cite{Brynjolfsson2004}
or  equation  (27) in
\cite{Brynjolfsson2009}.
The Hubble constant in the plasma theory  is  therefore
\begin{equation}
H_0 =3.077\,10^5 \neave  \h0units.
\label{h0plasma}
\end{equation}
A second mechanism  suggests a photo-absorption process
between the photon  and the electron: in this case, the
Hubble constant  is
\begin{equation}
H_0 = \frac{2 \, n_e  h r_e}{m_e}
\quad  ,
\end{equation}
where $n_e$ is the electron density,
      $h$   is the Planck constant,
      $r_e$ is the radius of the electron,
and   $m_e$ is the mass
of the electron, see  equation (2) in \cite{Ashmore2006}.
Once  the numerical values of the constants are
inserted in MKS,
the density of the electrons is substituted with
the averaged electron density in CGS,
we obtain
\begin{equation}
H_0=1.2649 \, 10^8 \neave  \h0units
\quad .
\label{h0photoabsortion}
\end{equation}
The investigation of the line
shift in dense and hot plasmas
can be found in
\cite{Nguyen1986,Leng1995,Saemann1999,Zhidkov1999,WangYang2007}.
As an example,
the experimental verification of the redshift of the
spectral line  of mercury as due to the surrounding
electrons      can be found in Figure  \ref{lineshift},
see also \cite{Ashmore2011}.
% figure  lineshift
% end figure  lineshift

\section{Two analytical  models}

\label{sectiontwo}

The two mechanisms for the redshift
here considered require the
evaluation of the averaged density
of electrons along the line of sight, which can be computed
by
\begin{equation}
\neave = \frac{\int_0^D n_e(x) dx}{D}
\quad  ,
\label{neaveint}
\end{equation}
where $D$ is the considered distance.
%modifica 
This will be  called the fundamental  integral.
%fine  modifica
We now present two models which are built in spherical  symmetry.

\subsection{The rim model}

\begin{figure*}
\begin{center}
\includegraphics[width=10cm]{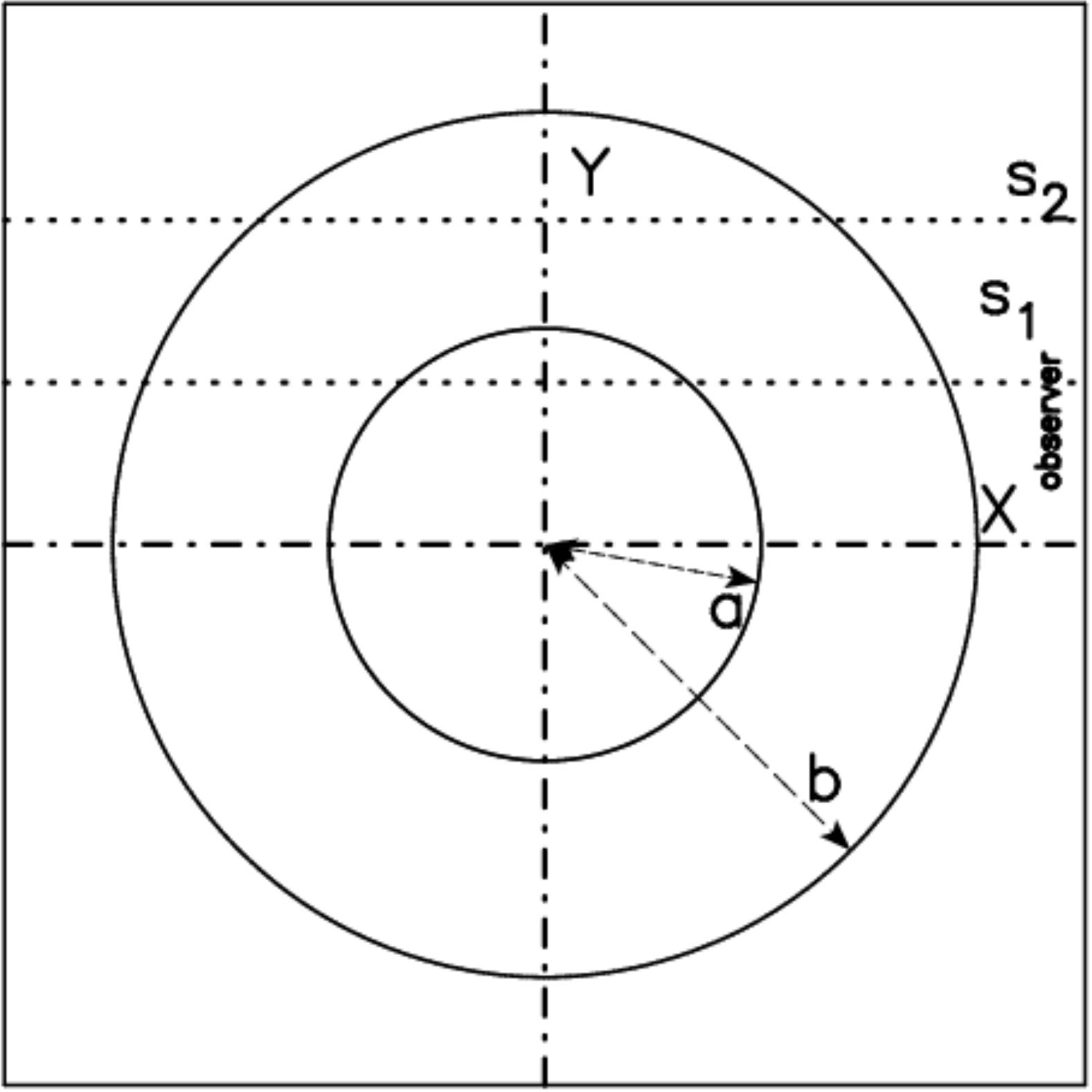}
\end {center}
\caption
{
The two circles (sections of spheres)  which
include the region
with an enhancement  in  electron  density
are   represented by
full lines.
The observer is situated along the $x$ direction, and
two lines of sight are indicated.
}
\label{configab}%label
    \end{figure*}

\begin{figure*}
\begin{center}
\includegraphics[width=10cm]{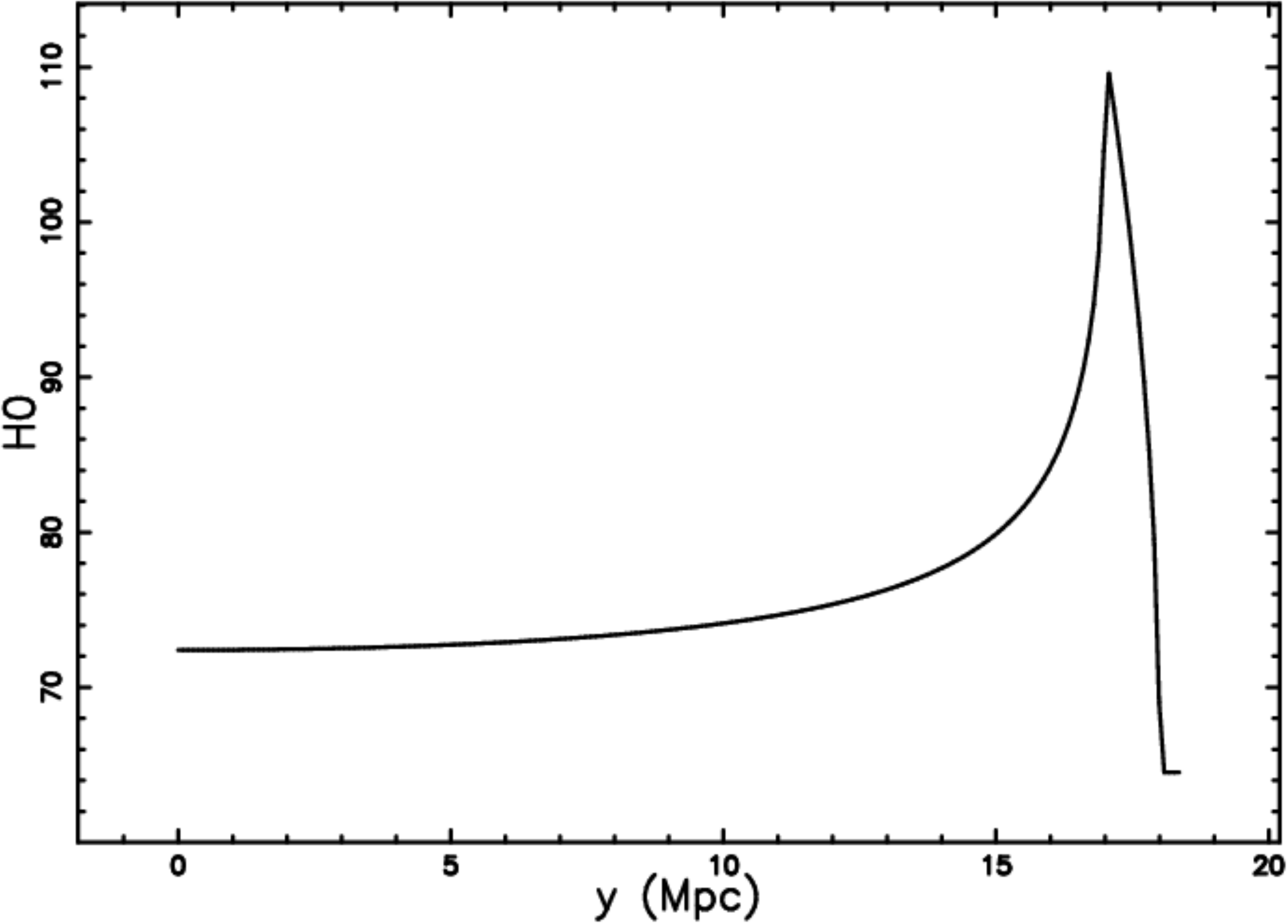}
\end {center}
\caption
{
Value of $H_0$
for the photo-absorption process
as function of the position;
$b=18$ Mpc,
$a=17$ Mpc,
$n_0= 5.1\,10^{-7} \neunits$,
and $n_1= 1.63\,10^{-6} \neunits$.
}
\label{h0ab}%label
    \end{figure*}

The radius of  the cosmic voids  as given  by
the Sloan Digital Sky Survey (SDSS)
R7,
has been modeled by spheres
which have averaged radius
$\bar{R}= \frac{18.23}{h}$\ Mpc,
see \cite{Vogeley2012}.
This means that the galaxies are situated on the  thick surface of spheres
having thickness $t$ much smaller than the averaged radius.
We now suggest that the density of the free electrons follows
the previous trend, and  therefore
\begin{eqnarray}
n_e(r) = n_0 & if &\quad \, 0 \,\le \,r \, <a   \nonumber \\
n_e(r) = n_1 & if &\quad \, a \,\le \,r \, <b   \\
n_e(r) = n_0 & if &\quad \, r  \ge b           \nonumber
\end{eqnarray}
where $r$ is the distance from the origin of a Cartesian 3D reference system.
This means that the electron density rises from $n_0$ at the center of the
sphere to $n_1$ at  $r=a$, remains constant up to $r=b$ (the radius of the void),
and then falls again to $n_0$ outside the sphere.
The fundamental integral of the density as
represented by Equation (\ref{neaveint})
can be done in the $x$-direction
over  the length $D=2b$
and is split into two parts,
\begin{eqnarray}
\mbox{part ~I,~three~pieces}  & \mbox{if} &\quad \, 0 \,\le \,y \, <a   \nonumber \\
\mbox{part~II,~two~pieces}   & \mbox{if} &\quad \, a \,\le  \,y  \, <b  \nonumber
\end{eqnarray}
corresponding to the lines of
sight $s_1$ and $s_2$ in
Figure~\ref{configab}.
% figure  configab
% end figure  configab
The result  of the integral  of the averaged
electron density, see (\ref{neaveint}),
is
\begin{eqnarray}
\neave=
\frac{
n_{{0}}b-n_{{0}}\sqrt {{b}^{2}-{y}^{2}}+n_{{1}}\sqrt {{b}^{2}-{y}^{2}}
-n_{{1}}\sqrt {{a}^{2}-{y}^{2}}+n_{{0}}\sqrt {{a}^{2}-{y}^{2}}
}{b}                              \nonumber \\
 \mbox{if} \quad \, 0 \,\le \,y \, <a   \nonumber \\
~                                          \\
\neave=\frac
{
n_{{0}}b-n_{{0}}\sqrt {{b}^{2}-{y}^{2}}+n_{{1}}\sqrt {{b}^{2}-{y}^{2}}
}
{
b
}
\nonumber  \\
 \mbox{if} \quad \, a \,\le  \,y  \, <b  \nonumber  \quad .
\label{neave}
\end{eqnarray}
The Hubble constant can be obtained  from the averaged
electron density by a multiplication with  a numerical
factor,
see Equation (\ref{neaveint}) for the
photo-absorption process.
Figure \ref{h0ab} reports the behavior
of Hubble's constant  as a function  of the
observer's  position
and Table \ref{dataab}, the statistical parameters
along the line of sight.
% figure  h0ab
% end figure  h0ab

\begin{table}[ht!]
\caption {
The Hubble constant
for the photo-absorption process.
}
\label{dataab}
\begin{center}
\begin{tabular}{|c|c|c|}
\hline
entity & definition & value   \\
\hline
$N$ & No. of samples    & 200 \\
$\bar {x}$ & average & 76.27           $\h0units$ \\
$\sigma $ & standard~deviation & 7.16b $\h0units$ \\
$H_0,max$ & maximum & 109.6            $\h0units$    \\
$H_0,min$ & minimum & 64.51            $\h0units$ \\
\hline
\end{tabular}
\end{center}
\end{table}

\subsection{The auto-gravitating  model}

We now present two solutions of two differential equations
which can represent the physical basis of two different PDFs.
On solving the fundamental hydrostatic equation
for the magneto-hydrostatic model in the direction perpendicular
to the galactic plane (the $x$-direction),
the solution for the density   is
\begin{equation}
\bar{\rho}(x) =\bar{\rho}(0) \exp{-(\frac{x}{H})^2)}
\quad ,
\label{solutionnormal}
\end{equation}
where $\bar{\rho}(x)$ is the mean gas density and
$H$ is the layer half-thickness,
see equation (6.142) in  \cite{Scheffler1987}.
The density profile of a thin
self-gravitating disk of stars
which is characterized by a
Maxwellian distribution in velocity
and  a distribution which varies
only in the $x$-direction is
\begin{equation}
n(x) = n_0 sech^2 (\frac{x}{2\,x_0})
\quad ,
\label{sech2}
\end{equation}
where $n_0$ is the star density at $x=0$,
$x_0$ is a scaling parameter,
and  sech is the hyperbolic secant,
see  formula (2.31) in (\cite{Padmanabhan_III_2002}).
The physical solution represented by (\ref{solutionnormal})
becomes, when normalized,  
the  Normal (Gaussian) distribution which
has PDF
\begin{equation}
N(x;\sigma) =
\frac {1} {\sigma (2 \pi)^{1/2}} \exp ({- {\frac {x^2}{2\sigma^2}}} )
\quad
-\infty < x < \infty
\quad .
\label{gaussian}
\end{equation}
The mean is
\begin{equation}
E(x;\sigma)=0
\quad,
\end{equation}
and  the variance is
\begin{equation}
VAR(x;\sigma) = \sigma^2
\quad .
\end{equation}

\begin{figure*}
\begin{center}
\includegraphics[width=10cm]{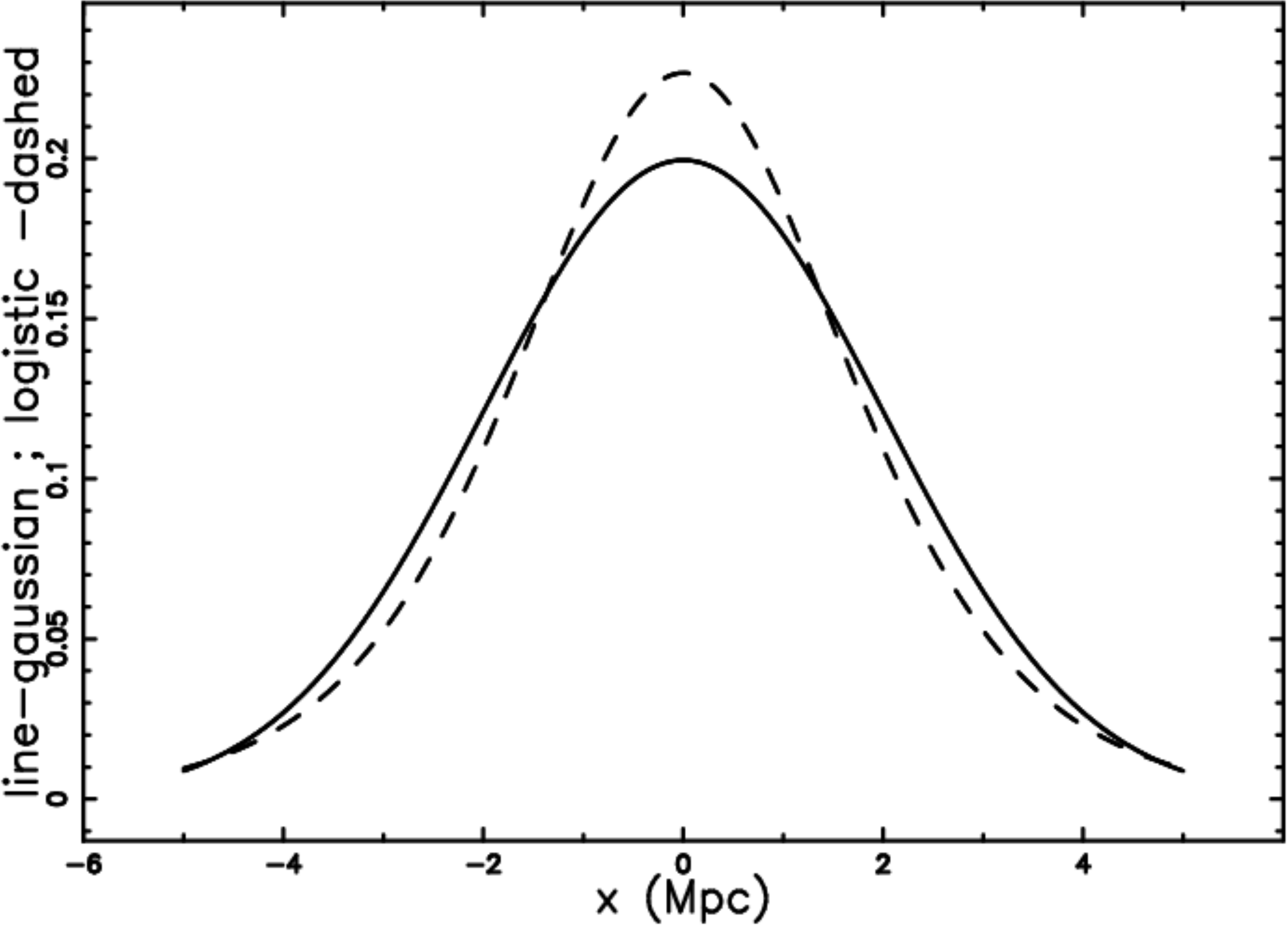}
\end {center}
\caption
{
Normal PDF (full line) and
logistic as
represented by Eq. (\ref {logistic})
(dashed line)
when $\sigma=2 { \mbox{Mpc}}/h $.
}
\label{sechgauss}
    \end{figure*}

\begin{figure*}
\begin{center}
\includegraphics[width=10cm]{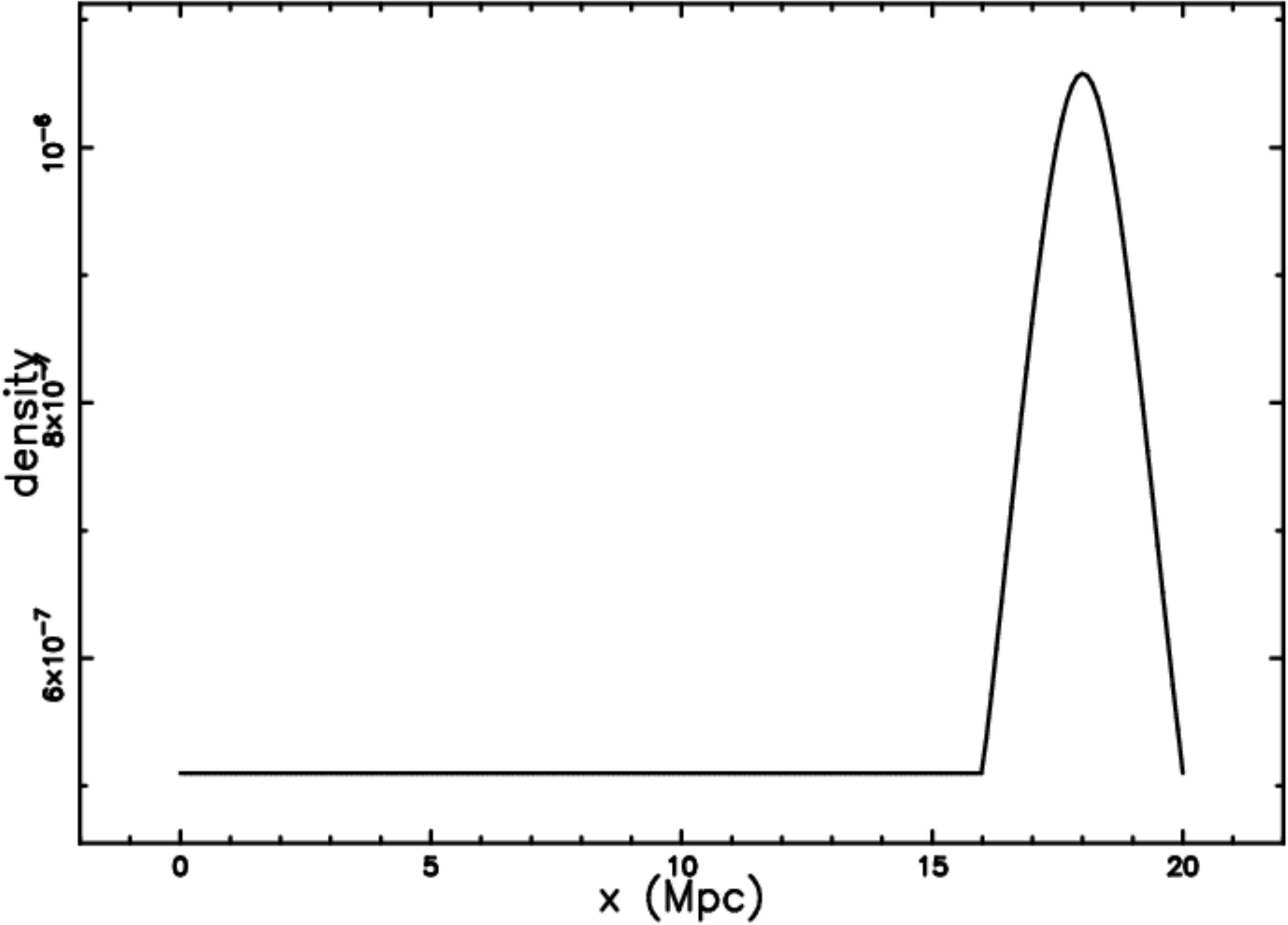}
\end {center}
\caption
{
Profile of the three-part density function of free electrons
as a function of the distance from the
center of the void;
$n_0=5\,10^{-7} \neunits$ , $n=2$, $b=18\,\mbox{Mpc}/h$,
$\sigma=2$\ Mpc/$h$.
}
\label{profilevoid}
    \end{figure*}

\begin{figure*}
\begin{center}
\includegraphics[width=10cm]{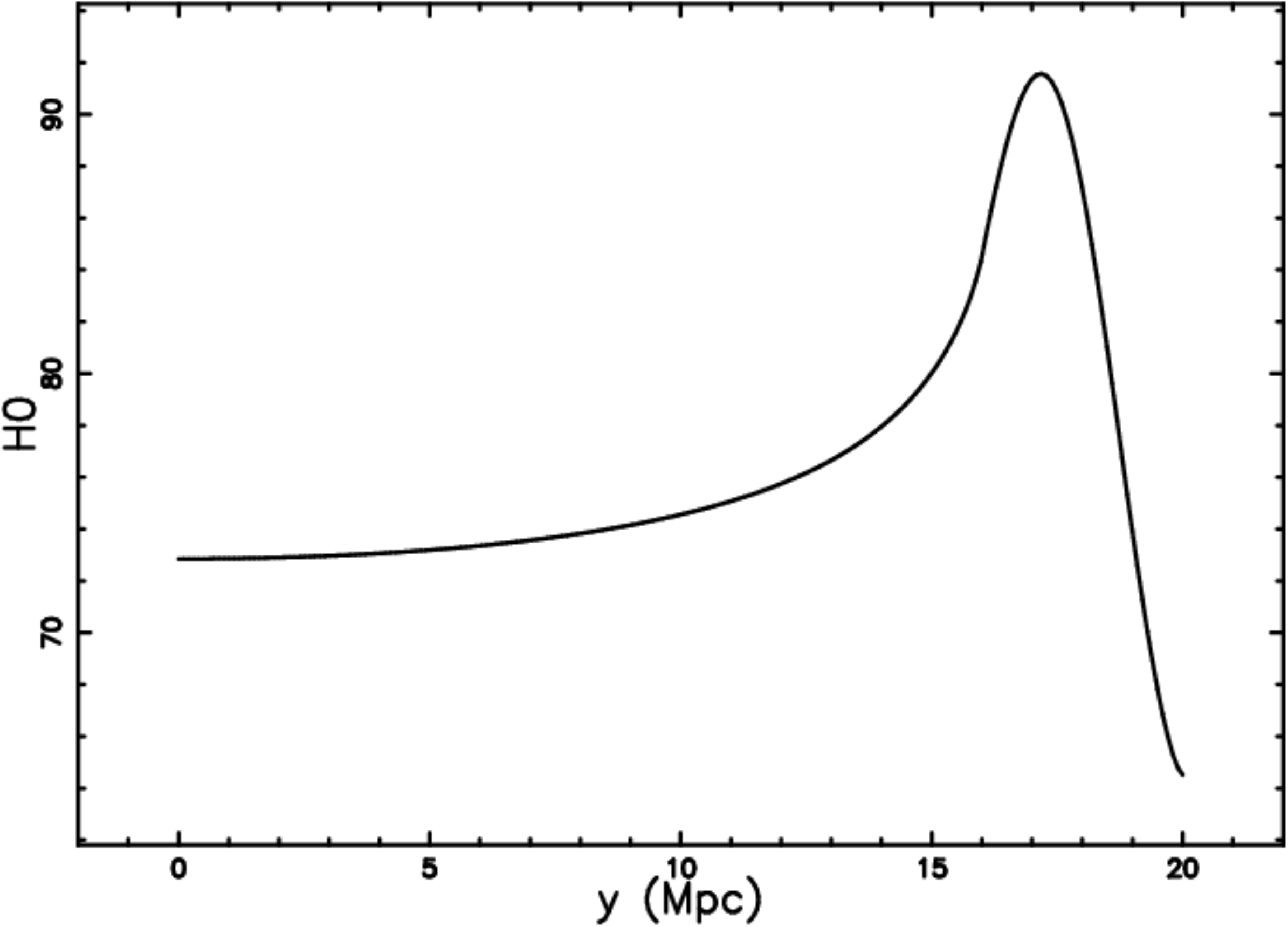}
\end {center}
\caption
{
Value of $H_0$
for the photo-absorption process
as a function of the position in
the three-part density of the free electrons;
$b=18$ Mpc,
$s=2 $ Mpc,
$n=1 $ Mpc, and
$n_0= 5.1\,10^{-7} \neunits$.
}
\label{h0abc}%label
    \end{figure*}

The physical law  represented by Equation (\ref{sech2})
can be converted  to a
probability density function (PDF),
the probability of
having a  given physical quantity at a distance
between $x$ and $x+dx$,
\begin{equation}
p(x;x_0)=
\frac{1}{4}\, \left( {\it sech} \left(\frac{ 1}{2}\,{\frac { \left| x \right| }{x_{{0}
}}} \right)  \right) ^{2}\frac{1}{x_0}
\quad .
\label{sech2prob}
\end{equation}
The range of existence of
this PDF
is in the interval $[-\infty, \infty]$.
The average value is
$E(x;x_0) =0$ and the variance is
\begin{equation}
VAR(x;x_0)=
\frac{1}{6}\,{x_{{0}}}^{2}{\pi }^{2}
\quad .
\end{equation}
This PDF can be transformed in such a way that
it can be compared with the
normal (Gaussian)
as  represented by Equation \ref{gaussian}.
The substitution $x_0={\frac {\sqrt {3}\sigma}{\pi }}$
transforms the PDF
(\ref{sech2prob})  into
\begin{equation}
p(x;\sigma)=
\frac{1}{12}\, \left( {\it sech} \left( \frac{1}{6}\,{\frac { \left| x \right| \pi \,
\sqrt {3}}{\sigma}} \right)  \right) ^{2}\pi \,\sqrt {3} \frac {1}{\sigma}
\quad .
\label{logistic}
\end{equation}
The average value is
$E(x;\sigma) =0$ and the variance is
\begin{equation}
VAR(x;\sigma)=\sigma^2
\quad .
\end{equation}
This PDF is referred to as the logistic function, see \cite{Balakrishnan1991handbook,univariate2,evans}.
The similarity with the normal distribution is
straightforward and
Figure \ref{sechgauss} reports
the two PDFs when  the value of
$\sigma$ is equal in both cases.
%begin figure sechgauss
%end   figure sechgauss
We now assume that the radial density  
of the free electrons
follows the logistic law as given by
Equation (\ref{sechgauss}), in  which $x$ is now
the distance from the surface of the sphere.
This assumption can be made when $\sigma \ll \bar {R}$.
In order to deal with distances from the spherical surfaces
greater than $\sigma$, where the density is supposed
to be constant, the  following three-part density function is
suggested.
\begin{eqnarray}
n_e(r) = n_0 & if &\quad \, 0 \,\le \,r \, <a   \nonumber \\
n_e(r) =n_{{0}} \frac
{
 \left( {\it sech} \left( 1/6\,{\frac { \left( r-b \right) \pi
\,\sqrt {3}}{\sigma}} \right)  \right) ^{2}
}
{
\left( {\it sech} \left(  0.9068\,n \right)  \right) ^{2}
}
& if &\quad \, a \,\le \,r \, < c   \\
n_e(r) = n_0 & if &\quad \, r  \ge c.           \nonumber
\label{profilelogistic}
\end{eqnarray}
Here, $r$ is the distance from the origin
of a Cartesian 3D reference system.
In order  to avoid   an  increase  in the number
of parameters,
$a$ and $c$ can  be parametrized  with the distance
$n\, \sigma$  after which the density is constant,
\begin {equation}
a = b- n \sigma  \qquad     c =b+n \sigma
\quad ,
\label{eqnac}
\end{equation}
where $n$  is an integer.
A simple evaluation says that the maximum density
is reached at $r=b$,
where the density
is
\begin{equation}
n_e(r) =
{\frac {{\it n_0}}{ \left( {\it sech} \left(  0.9068\,n \right)
 \right) ^{2}}}
\quad .
\end{equation}
This profile  density , see Equation (\ref{profilelogistic}),
as a function of the distance from the center of the void,
is shown in Figure~\ref{profilevoid}.
%begin figure profilevoid
%end   figure profilevoid
We insert, into the three-part density profile of the free electrons,
$r=\sqrt{x^2+y^2}$
and perform the integration
%modifica  
of (\ref {neaveint})
%finemodifica 
over $x$  maintaining
$y$ constant.
An   analytical  expression for the integral does not exist,
and the numerical  integration,
(using Boole's rule, see \cite{Abramowitz1965},
gives
the averaged density.
Under the  hypothesis of
the photo-absorption process, see Equation (\ref{neaveint}),
Figure \ref{h0abc} presents the behavior
of Hubble's constant  as a function  of the
observer's  position
and Table \ref{dataabc}, the statistical parameters
along the line of sight.

% figure  h0abc
% end figure  h0abc

\begin{table}[ht!]
\caption {
The Hubble constant
for the photo-absorption process
with
a three-part density of free electrons.
}
\label{dataabc}
\begin{center}
\begin{tabular}{|c|c|c|}
\hline
entity & definition & value   \\
\hline
$N$ & No. of samples    & 200 \\
$\bar {x}$ & average & 76.25           $\h0units$ \\
$\sigma $ & standard~deviation & 5.48  $\h0units$ \\
$H_0,max$ & maximum & 91.56            $\h0units$    \\
$H_0,min$ & minimum & 64.51            $\h0units$ \\
\hline
\end{tabular}
\end{center}
\end{table}

\section{A theoretical  model}

\label{sectiontheoretical}
\begin{figure*}
\begin{center}
\includegraphics[width=10cm]{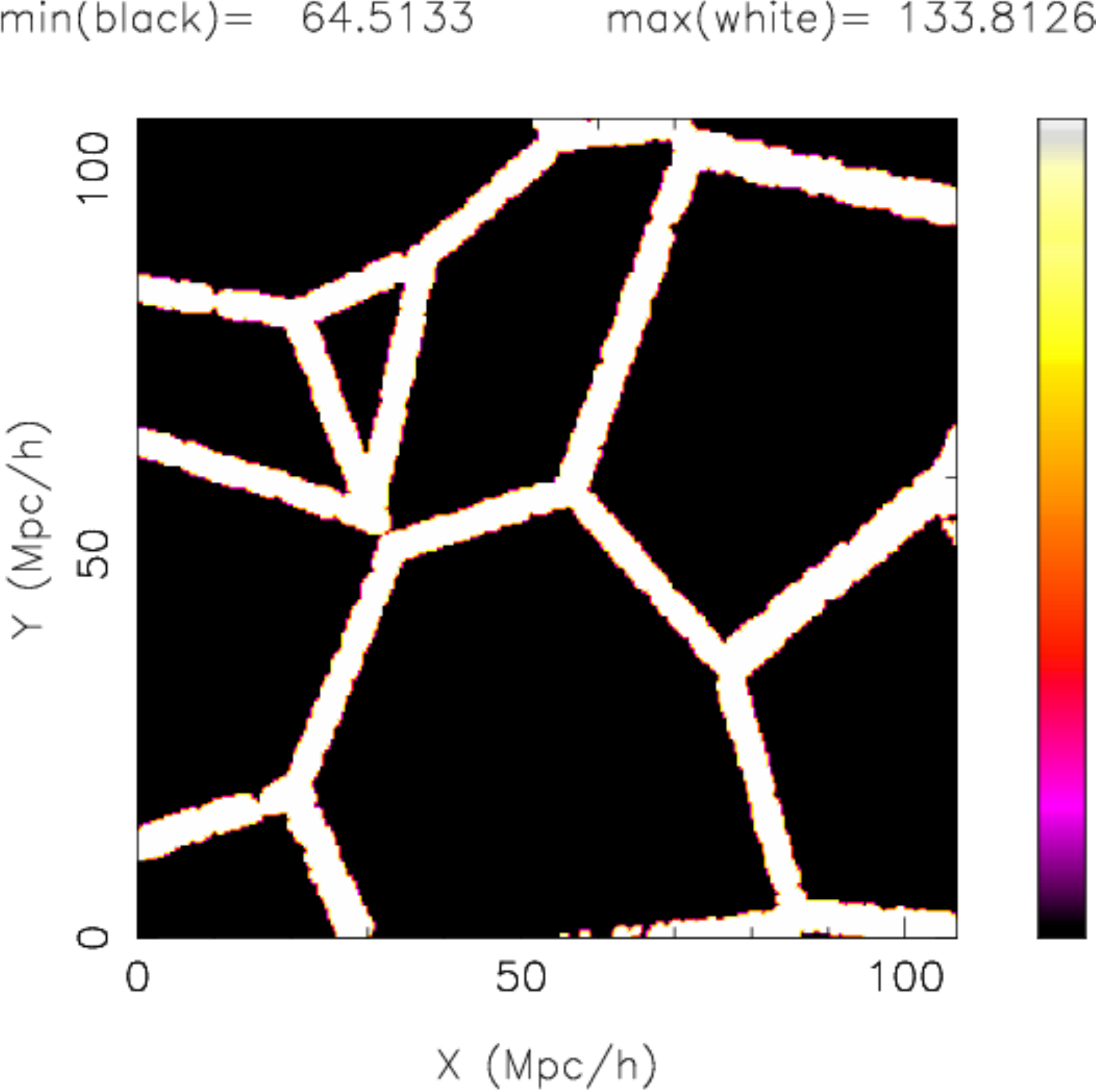}
\end {center}
\caption
{
Contour plot of $H_0$
for the photo-absorption process
as a function of the position in a 2D cut;
$\sigma=1.9$\ Mpc,
$n_0= 5.1\,10^{-7} \neunits$ and
$n=1$.
}
\label{clusters_h0}%label
    \end{figure*}

A possible model for the spatial distribution of galaxies
is represented by the Voronoi diagrams,
but two requirements should be satisfied.
The first is that
the average radius of the voids be
$<R>=18.23 h^{-1}$\ Mpc,
which is the effective radius in  SDSS DR7, see Table 6
in \cite{Zaninetti2012e}.
The second requirement
is connected to
the fact that
the effective  radius  of the cosmic
voids as  deduced from
catalog SDSS R7  is  represented  by a Kiang function
with $c \approx 2$.
This  means that  we are considering
a non-Poissonian Voronoi
Tessellation (NPVT).
The density of free electrons can be found:
(i) computing the distance
$d$ of a 3D grid point from the nearest face,
(ii) inserting such a distance in the following
two-part density
\begin{eqnarray}
n_e(d) =n_{{0}} \frac
{
 \left( {\it sech} \left( 1/6\,{\frac {  d  \pi
\,\sqrt {3}}{\sigma}} \right)  \right) ^{2}
}
{
\left( {\it sech} \left(  0.9068\,n \right)  \right) ^{2}
}
& if &\quad \,   d \, < c   \nonumber \\
                                      \\
n_e(d) = n_0 & if &\quad \, d  \ge c           \nonumber
\label{profileface}
\end{eqnarray}
with $c=n\,\sigma$.
Given a cubic box of size  100 Mpc,
Figure \ref{clusters_h0}
shows the contour plots of the Hubble constant
when a plane  crossing the center is considered,
and  Table \ref{datavoronoi} gives the statistics.
%siamoqui
In this case, the fundamental integral of the density as
represented by Equation (\ref{neaveint}) is evaluated 
numerically along lines belonging to the selected 
plane.
% figure  clusters_h0
% end figure  clusters_h0

\begin{figure*}
\begin{center}
\includegraphics[width=10cm]{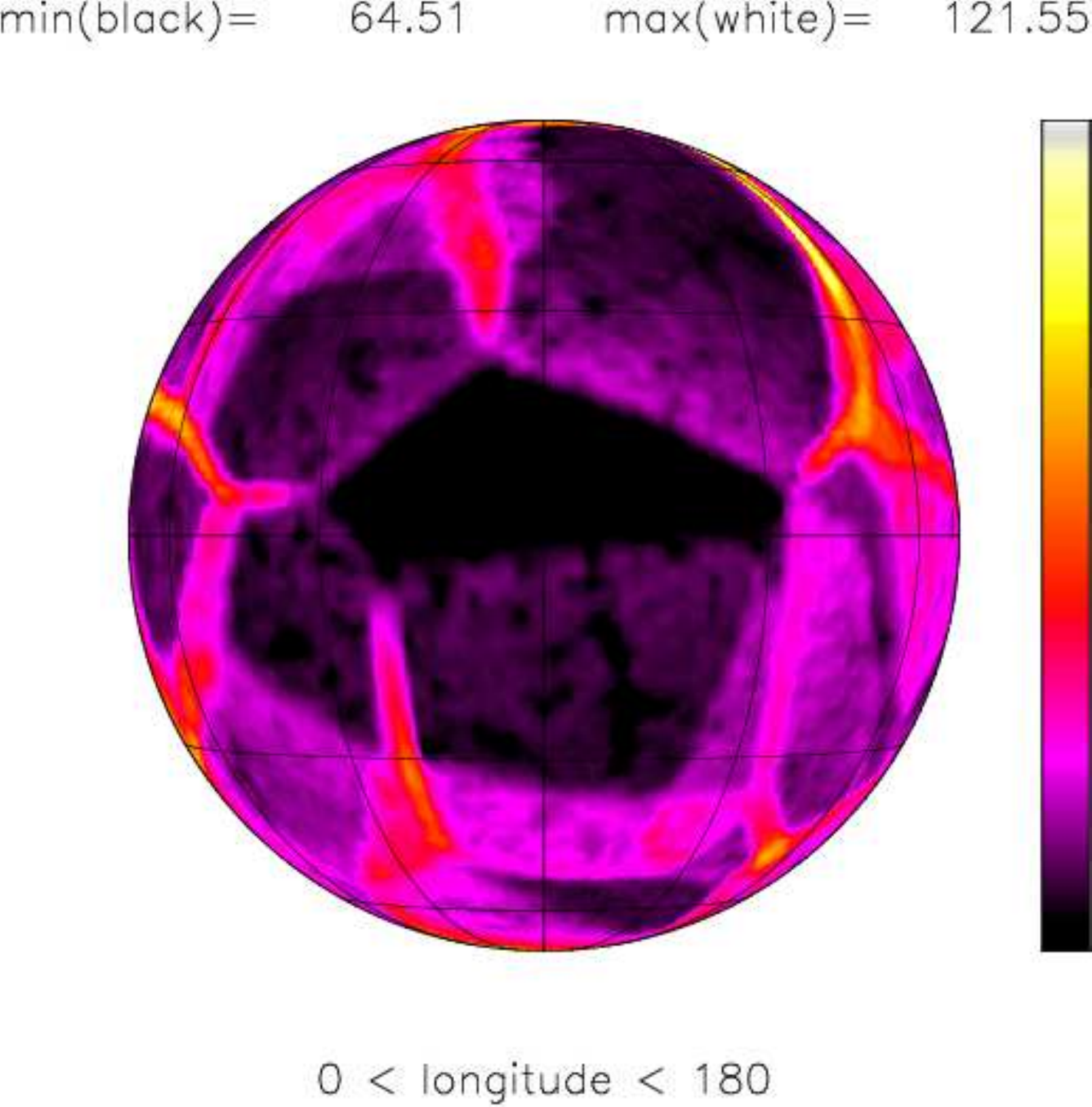}
\end {center}
\caption
{
Contour plot of $H_0$
on the surface of a sphere,
parameters as in Figure \ref{clusters_h0}.
}
\label{h0globe1}%label
    \end{figure*}

\begin{table}[ht!]
\caption {
The Hubble constant
for the photo-absorption process
in
the 2D cut of a Voronoi diagram.
}
\label{datavoronoi}
\begin{center}
\begin{tabular}{|c|c|c|}
\hline
entity & definition & value   \\
\hline
$N$ & No. of samples    & 301 \\
$\bar {x}$ & average & 76.45           $\h0units$ \\
$\sigma $ & standard~deviation & 6.11  $\h0units$ \\
$H_0,max$ & maximum &   96.51          $\h0units$ \\
$H_0,min$ & minimum &   68.88          $\h0units$ \\
\hline
\end{tabular}
\end{center}
\end{table}
The fundamental integral  (\ref{neaveint})
which gives the averaged
density can be done also along an arbitrary
line characterized by a given galactic latitude and
longitude.
Figure \ref{h0globe1} shows  the  contours of the
Hubble constant  on the surface of a sphere
using the the subroutine  GLOBE which belongs to the
package PGXTAL.

% figure  h0globe1
% end figure  h0globe1

\section{The astronomer's model}

\label{sectionastronomers}
\begin{figure}
\begin{center}
\includegraphics[width=10cm]{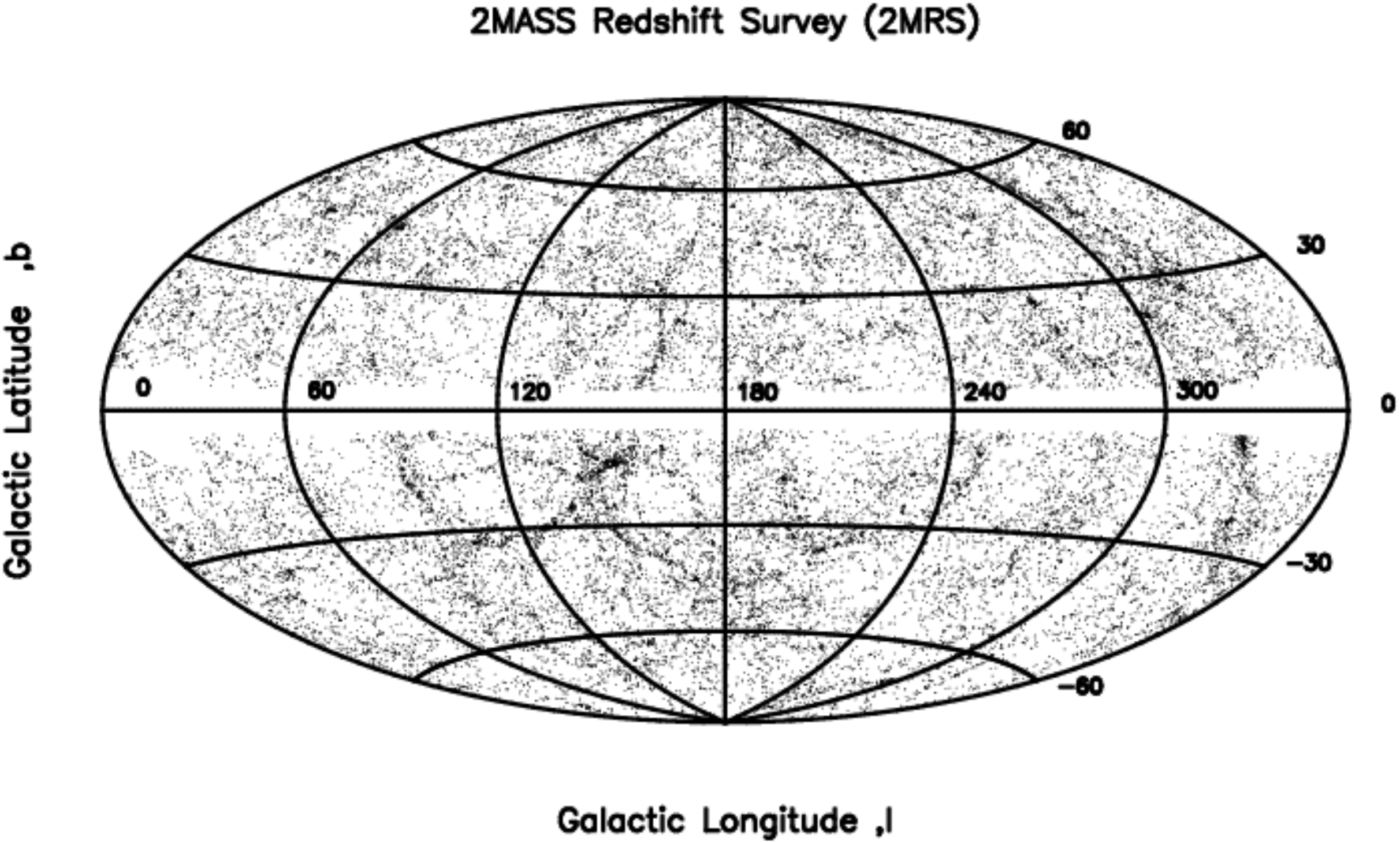}
\end {center}
\caption
{
Hammer--Aitoff  projection in galactic coordinates
of  the 2MRS  galaxies  when
$z \leq   0.04$.
}
          \label{2mrsaitof}%
    \end{figure}

We  analyze  the 3D spatial  distribution of
galaxies given by
the 2MASS Redshift Survey (2MRS),
see Figure \ref{2mrsaitof}.
%begin figure 2mrsaitof
% end figure 2mrsaitof
An updated  real 3D distribution of galaxies
as given by the 2MRS catalog is the
first
ingredient of the astronomer's model.
The self-gravitating sphere of polytropic
gas is regulated
by the Lane--Emden differential equation
of the second order,
\begin{equation}
{\frac {d^{2}}{d{x}^{2}}}Y \left( x \right) +2\,{\frac {{\frac {d}{dx}
}Y \left( x \right) }{x}}+ \left( Y \left( x \right)  \right) ^{n}=0
\quad ,
\end{equation}
where $n$  is an integer,
see
\cite{Lane1870,Emden1907,Chandrasekhar_1967,Binney1987,Zwillinger1989}.

The solution  $Y \left( x \right)_n$
produces the density profile
\begin{equation}
\rho = \rho_c Y \left( x \right)_n^n
\quad ,
\end{equation}
where $\rho_c$ is the density at $x=0$.

Analytical solutions exist for $n=0,1$, and 5.
The analytical  solution for $n=0$ is
\begin{equation}
Y(x) = \frac{sin(x)}{x}
\quad,
\end{equation}
and has therefore an oscillatory behavior.
The analytical  solution for $n=5$ is
\begin{equation}
Y(x) ={\frac {1}{{(1+ \frac{{x}^{2}}{3})^{1/2}}} }
\quad ,
\end{equation}
and the density for $n=5$
is
\begin{equation}
\rho(x) =\rho_c {\frac {1}{{(1+ \frac{{x}^{2}}{3})^{5/2}}} }
\label{densita5}
\quad .
\end{equation}
The asymptotic  behavior for large values of $x$
of  the density when  $n=5$ is
\begin{equation}
\rho(x) =
9\,{\frac {\sqrt {3}}{{x}^{5}}}+O \left( \frac{1}{x^7} \right)
\quad .
\end{equation}
On introducing the scale , $b$, and normalizing the
function (\ref{densita5}), we obtain the
Lane--Emden5 (i.e., $n=5$)  PDF, LE5,
\begin{equation}
LE5(x;b) =\frac
{
\sqrt {3}
}
{
2\, \left( 1+\frac{1}{3}\,{\frac {{y}^{2}}{{b}^{2}}} \right) ^{5/2}b
}
\quad ,
\end{equation}
which has average value
\begin{equation}
E(x;b)=
\frac{1}{2}\,\sqrt {3}b
\quad ,
\end{equation}
and variance
\begin{equation}
VAR(x;b)=\frac{3}{4}\,{b}^{2}
\quad .
\end{equation}
The density $\rho$  can be written as
$\rho(x) = n(x) m_H $ where   $m_H$ is the mass of hydrogen.
On assuming the neutrality of the charges,
the number density of the electrons is
\begin{equation}
n_e(x) = n_c  LE5(x;b)
\quad ,
\end{equation}
where $n_c$ is the  central
number density of the electrons.
We now assume that
the Lane--Emden profile can be applied
from $x=0$ to  a given value  of $x$
after which the number density is decreased
by a given factor $f$.
After this length,  the  number density of electrons
has a constant value $n(x)=n_0$, where
$n_0$ is a constant.
The inequality  which  fixes the previous
assumption is
\begin{equation}
\cases{
n_e(x)=f{\it n_0} \left( 1+1/3\,{\frac {{x}^{2}}{{b}^{2}}} \right)
^{-5/2}&when \, $0<x<\sqrt {-3+3\,{f}^{2/5}}b$\cr
{\it n_e(x)= n_0}& when \, $x> \sqrt
{-3+3\,{f}^{2/5}}$.\cr}
\end{equation}

\begin{figure*}
\begin{center}
\includegraphics[width=10cm]{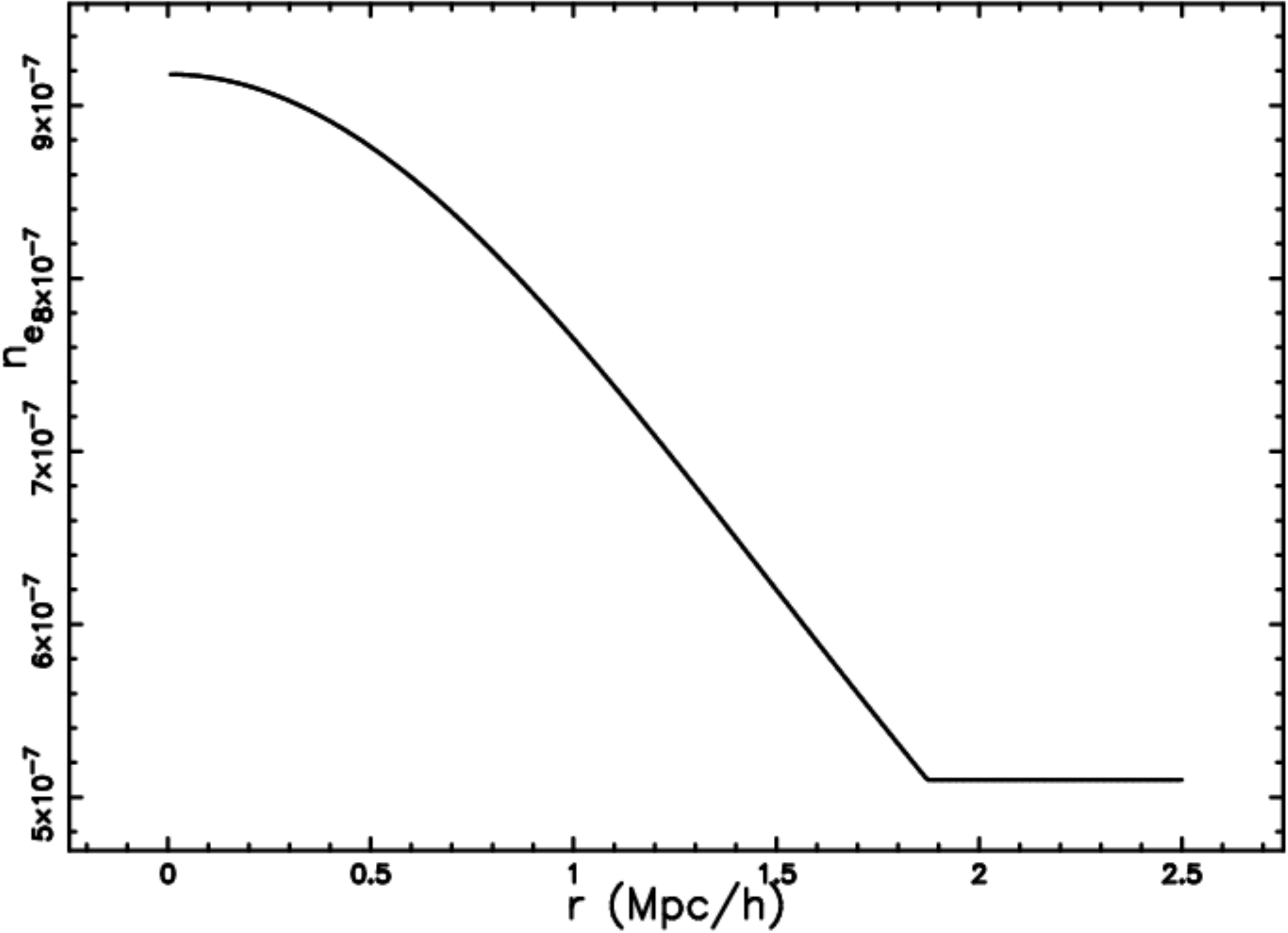}
\end {center}
\caption
{
Profile of  density of free electrons
as a function of the distance from the
center of the void;
$n_0=5.1 \,10^{-7} \neunits$ , $b=2.55$\ Mpc/$h$,
$f= 2.35$ .
}
\label{emdenprofile}
    \end{figure*}

\begin{figure*}
\begin{center}
\includegraphics[width=10cm]{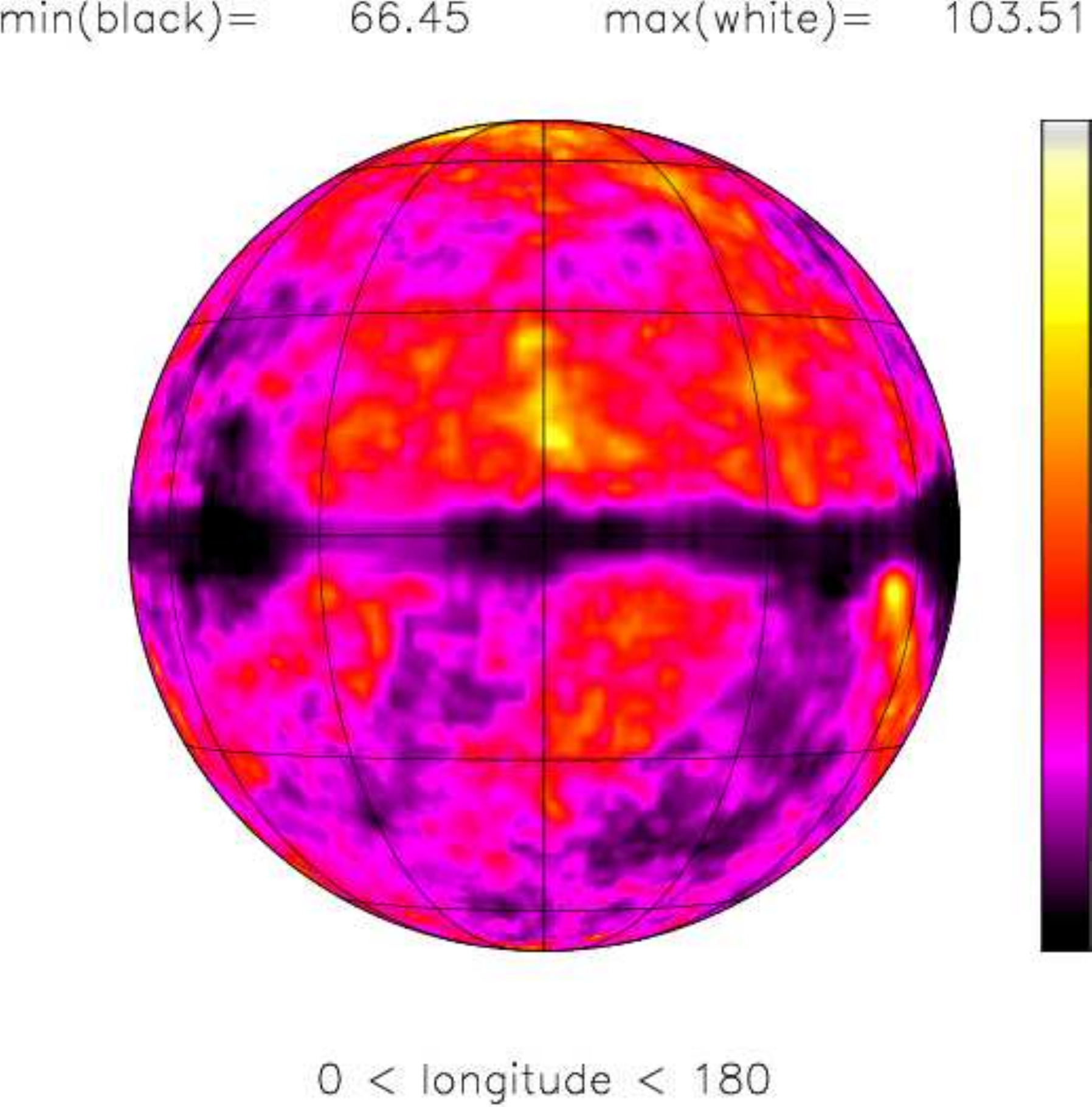}
\end {center}
\caption
{
Contour plot of $H_0$  as given by
the astronomer's model
on the surface of a sphere,
parameters as in Figure
\ref{emdenprofile}.
Galactic  longitude   between
0$^{\circ}$ and  180$^{\circ}$.
}
\label{2mrshg1}%label
    \end{figure*}

\begin{figure*}
\begin{center}
\includegraphics[width=10cm]{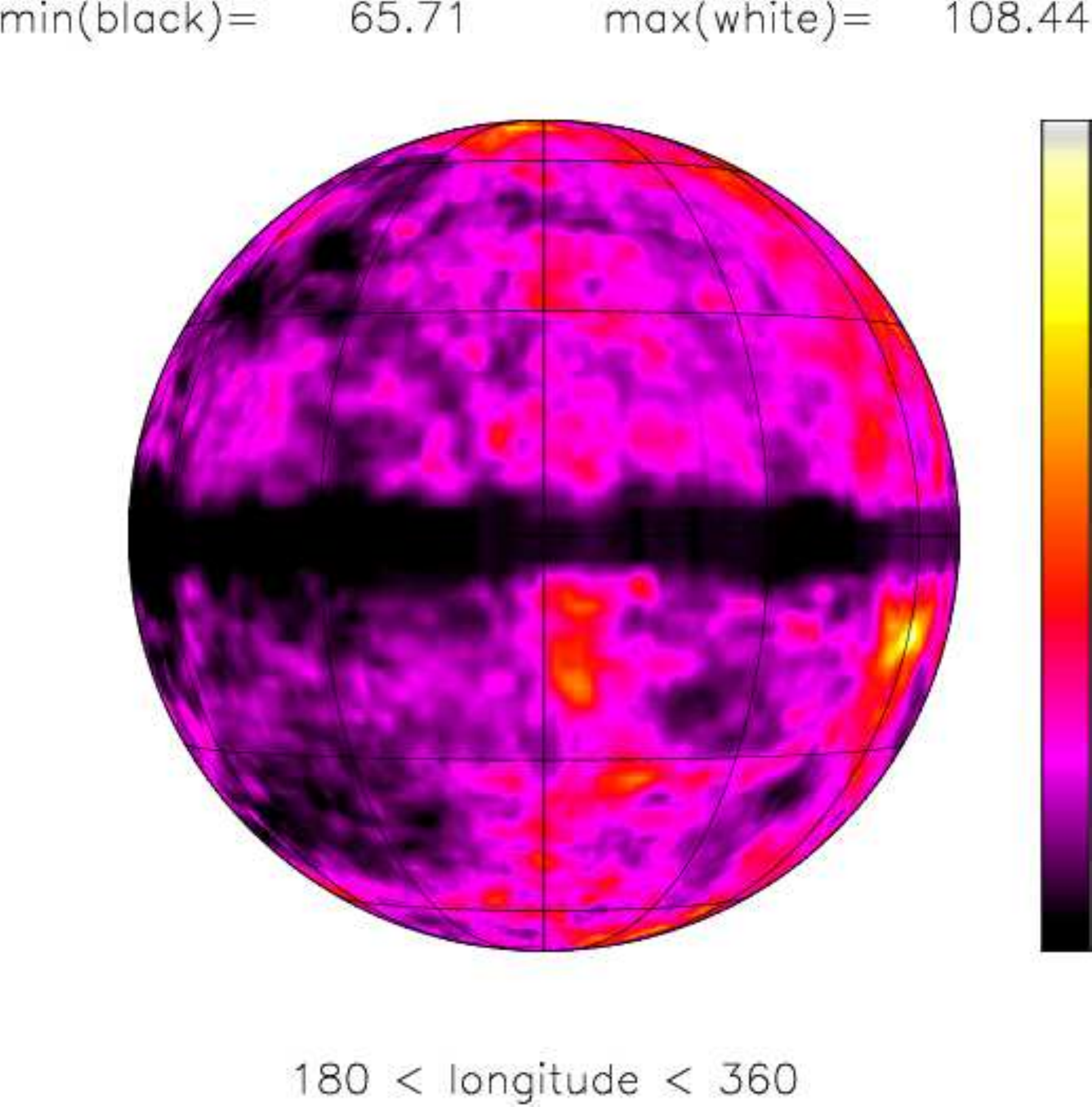}
\end {center}
\caption
{
Contour plot of $H_0$  as given by
the astronomer's model
on the surface of a sphere,
parameters as in Figure
\ref{emdenprofile}.
Galactic  longitude   between 180$^{\circ}$ and  360$^{\circ}$.
}
\label{2mrshg2}%label
    \end{figure*}

\begin{figure*}
\begin{center}
%richiede molti pixels
\includegraphics[width=10cm]{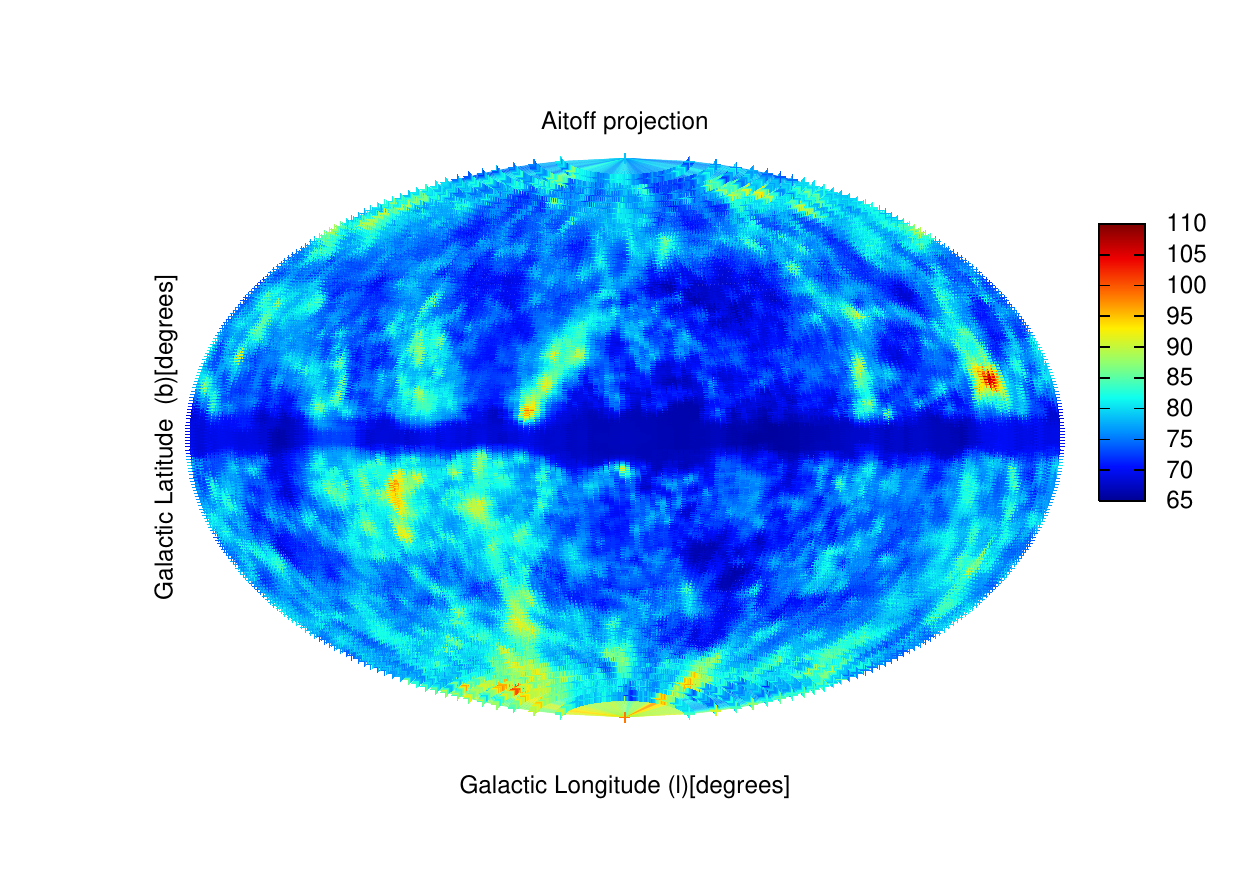}
\end {center}
\caption
{
False color contour plot of $H_0$  as given by
the astronomer's model in the
Aitoff--Hammer equal-area map in galactic coordinates
with the Galactic
center in the middle,
parameters as in Figure
\ref{emdenprofile}.
}
\label{h0aitof}%label
    \end{figure*}

A typical example with the parameters
that  will be used in the
forthcoming  simulation is given in Figure
\ref{emdenprofile}.
%begin figure emdenprofile
%end   figure emdenprofile
The radial distribution  of electrons
as given by the
Lane--Emden ($n=5$) profile
is the second
ingredient of the astronomer's model.
Figures
\ref{2mrshg1}, respectively, 
\ref{2mrshg2},
give  the  contours of the
Hubble constant  on the surface of the
first, respectively, second, half-sky
where the fundamental integral  (\ref{neaveint})
which gives the averaged
density has been used.
The contours of the
Hubble constant can also be drawn
in    the Aitoff--Hammer  projection,
see   Figure \ref{h0aitof},
and
Table \ref{dataastronomers} gives
the statistics of the Hubble constant.
% figure  2mrshg1
% end figure  2mrshg1

% figure  2mrshg2
% end figure  2mrshg2

% figure h0aitof
% end figure h0aitof

\begin{table}[ht!]
\caption {
The Hubble constant
for the photo-absorption process
in
the astronomer's model.
}
\label{dataastronomers}
\begin{center}
\begin{tabular}{|c|c|c|}
\hline
entity & definition & value   \\
\hline
$n$ & No. of samples    & 16471                      \\
$\bar {x}$ & average & 76.1                  $\h0units$ \\
$\sigma $ & standard~deviation &   5.5       $\h0units$ \\
$H_0,max$ & maximum &              105.91    $\h0units$ \\
$H_0,min$ & minimum &              65.69     $\h0units$ \\
\hline
\end{tabular}
\end{center}
\end{table}

\section{Conclusions}

The fact that the recent observations of galaxies
reveal a cellular structure as opposed 
to a homogeneous  structure,
see Section \ref{nonhomo}, 
allows exploring  a definition of the Hubble constant
outside of the standard cosmology.
We therefore explained
the anisotropies
in the Hubble constant
by variations  in the number
of electrons on the line of sight  in the framework of
the photo-absorption process.
The observed average value and variance
in the all sky measures
have been reproduced  in the analytical, simulated,
and astronomer's models,
see Tables
\ref{dataab},
\ref{dataabc},
\ref {datavoronoi},
and  \ref{dataastronomers}.
The statistical deviations
of the Hubble constant
are high  and the standard
uncertainty is  10.57 $\h0units$,
which means a  relative  standard
uncertainty of 0.13, see Table \ref{hubblemanyhst}.
We have  reproduced this
high   statistical fluctuation  and
the relative  standard
uncertainty is
0.0938 and 0.0718
for the two analytical models,   see
Tables
\ref{dataab} and
\ref{dataabc},
0.079   for  the simulated model, see Table
\ref {datavoronoi},
and   0.072  for  the astronomer's model,
see Table \ref{dataastronomers}.
% da sviluppare  in futuro
%giant radio sources which have linear size around 0.3 Mpc/h,
%%ee  \cite{Machalski2006}

\section*{Acknowledgments}
We thank
D. S. Sivia,  who  has
written the  set of routines
PGXTAL, for 3D plotting structures and density
maps.

%\bibliography{biblio}

\begin{thebibliography}{10}
\expandafter\ifx\csname url\endcsname\relax
  \def\url#1{\texttt{#1}}\fi
\expandafter\ifx\csname urlprefix\endcsname\relax\def\urlprefix{URL }\fi

\bibitem{Hubble1929}
E.~{Hubble}, {A Relation between Distance and Radial Velocity among
  Extra-Galactic Nebulae}, {\it Proceedings of the National Academy of Science}
  {\bf 15} (1929), 168--173.

\bibitem{Freedman2012}
W.~L. {Freedman}, B.~F. {Madore}, V.~{Scowcroft}, C.~{Burns}, A.~{Monson},
  S.~E. {Persson}, M.~{Seibert}, J.~{Rigby}, {Carnegie Hubble Program: A
  Mid-infrared Calibration of the Hubble Constant}, {\it \apj} {\bf 758}
  (2012), 24.

\bibitem{CODATA2012}
P.~J. {Mohr}, B.~N. {Taylor}, D.~B. {Newell}, {CODATA recommended values of the
  fundamental physical constants: 2010}, {\it Reviews of Modern Physics} {\bf
  84} (2012), 1527--1605.

\bibitem{Friedmann1922}
A.~{Friedmann}, {{\"U}ber die Kr{\"u}mmung des Raumes}, {\it Zeitschrift fur
  Physik} {\bf 10} (1922), 377--386.

\bibitem{Friedmann1924}
A.~{Friedmann}, {\"U}ber die m{\"o}glichkeit einer welt mit konstanter
  negativer kr{\"u}mmung des raumes, {\it Zeitschrift f{\"u}r Physik A Hadrons
  and Nuclei} {\bf 21}~(1) (1924), 326--332.

\bibitem{Riess1998}
A.~G. {Riess}, A.~V. {Filippenko}, P.~{Challis}, A.~{Clocchiatti},
  {Observational Evidence from Supernovae for an Accelerating Universe and a
  Cosmological Constant}, {\it \aj} {\bf 116} (1998), 1009--1038.

\bibitem{Perlmutter1999}
S.~{Perlmutter}, G.~{Aldering}, G.~{Goldhaber}, R.~A. {Knop}, {Measurements of
  Omega and Lambda from 42 High-Redshift Supernovae}, {\it \apj} {\bf 517}
  (1999), 565--586.

\bibitem{Zwicky1929}
F.~{Zwicky}, {On the Red Shift of Spectral Lines through Interstellar Space},
  {\it Proceedings of the National Academy of Science} {\bf 15} (1929),
  773--779.

\bibitem{Brynjolfsson2004}
A.~{Brynjolfsson}, {Redshift of photons penetrating a hot plasma}, {\it
  arXiv:astro-ph/0401420} .

\bibitem{Brynjolfsson2009}
A.~{Brynjolfsson}, {Plasma-Redshift Cosmology: A Review}, in: {F.~Potter}
  (Ed.), \textit{Astronomical Society of the Pacific Conference Series}, Vol. 413,
  2009, 169--189.

\bibitem{Ashmore2006}
L.~{Ashmore}, Recoil between photons and electrons leading to the Hubble
  constant and CMB, {\it Galilean Electrodynamics} {\bf 17}~(3) (2006), 53.

\bibitem{Crawford2006}
D.~F. {Crawford}, {\it {Curvature Cosmology}}, Brown Walker, Boca Raton, FL, USA,
  2006.

\bibitem{Crawford2011}
D.~F. {Crawford}, {Observational Evidence Favors a Static Universe (Part I)},
  {\it Journal of Cosmology} {\bf 13} (2011), 3875--3946.

\bibitem{Mamas2010}
D.~L. {Mamas}, {An explanation for the cosmological redshift}, {\it Physics
  Essays} {\bf 23} (2010), 326.

\bibitem{Michelini2013}
M.~{Michelini}, The new cosmology rising from the quantum pushing gravity
  interaction---The case of accelerating universe, {\it Applied Physics Research}
  {\bf 5}~(5) (2013), 67--84.

\bibitem{Marmet2009}
L.~{Marmet}, {Survey of Redshift Relationships for the Proposed Mechanisms at
  the 2\^{}$\{$nd$\}$ Crisis in Cosmology Conference}, in: {F.~Potter} (Ed.),
  \textit{Astronomical Society of the Pacific Conference Series}, Vol. 413,
  2009, 315--335.

\bibitem{Karoji1975}
H.~{Karoji}, M.~{Moles}, {Markarian galaxies and the angular anisotropy of the
  Hubble `constant'}, {\it Academie des Sciences Paris Comptes Rendus Serie B
  Sciences Physiques} {\bf 280} (1975), 609--612.

\bibitem{Guthrie1976}
B.~N.~G. {Guthrie}, {Anisotropy in the Hubble Relation for the Brightest
  Galaxies in Clusters}, {\it \apss} {\bf 43} (1976), 425--431.

\bibitem{Fennelly1977}
A.~J. {Fennelly}, {Anisotropy in the Hubble parameter and large-scale
  cosmological inhomogeneity}, {\it \mnras} {\bf 181} (1977), 121--130.

\bibitem{Wiltshire2013}
D.~L. {Wiltshire}, P.~R. {Smale}, T.~{Mattsson}, R.~{Watkins}, {Hubble flow
  variance and the cosmic rest frame}, {\it \prd} {\bf 88}~(8) (2013), 083529.

\bibitem{Einstein1916}
A.~{Einstein}, {Die Grundlage der allgemeinen Relativit\"{a}tstheorie}, {\it
  Annalen der Physik} {\bf 354} (1916), 769--822.

\bibitem{Einstein1917}
A.~{Einstein}, {Kosmologische Betrachtungen zur allgemeinen
  Relativit\"{a}tstheorie}, {\it Sitzungsberichte der K\"{o}niglich Preu{\ss}ischen
  Akademie der Wissenschaften} (Berlin), (1917), 142--152.

\bibitem{Einstein1922}
A.~{Einstein}, Bemerkung zu der Franz Seletyschen arbeit---Beitr{\"a}ge zum
  kosmologischen system, {\it Annalen der Physik} {\bf 374} (1922),
  436--438.

\bibitem{Wang2006}
F.~{Wang}, {\it Physics with MAPLE: The Computer Algebra Resource for
  Mathematical Methods in Physics}, Wiley-VCH, New York, 2006.

\bibitem{McClure2007}
M.~L. {McClure}, C.~C. {Dyer}, {Anisotropy in the Hubble constant as observed
  in the HST extragalactic distance scale key project results}, {\it \na} {\bf
  12} (2007), 533--543.

\bibitem{Leo1994}
W. R. Leo, {\it Techniques for Nuclear and Particle Physics Experiments},
  Springer-Verlag, Berlin, 1994.

\bibitem{Zaninetti2010c}
L.~{Zaninetti}, {New formulas for the Hubble constant in a Euclidean static
  universe}, {\it Physics Essays} {\bf 23} (2010), 298--307.

\bibitem{Marmet1988}
P.~Marmet, A new non-Doppler redshift, {\it Physics Essays} {\bf 1}~(1) (1988),
  24--32.

\bibitem{Nguyen1986}
H.~{Nguyen}, M.~{Koenig}, D.~{Benredjem}, M.~{Caby}, G.~{Coulaud}, {Atomic
  structure and polarization line shift in dense and hot plasmas}, {\it \pra}
  {\bf 33} (1986), 1279--1290.

\bibitem{Leng1995}
Y.~{Leng}, J.~{Goldhar}, H.~R. {Griem}, R.~W. {Lee}, {C vi Lyman line profiles
  from 10-ps KrF-laser-produced plasmas}, {\it \pre} {\bf 52} (1995),
  4328--4337.

\bibitem{Saemann1999}
A.~{Saemann}, K.~{Eidmann}, I.~E. {Golovkin}, Isochoric heating of solid
  aluminum by ultrashort laser pulses focused on a tamped target, {\it Physical
  Review Letters} {\bf 82}~(24) (1999), 4843--4846.

\bibitem{Zhidkov1999}
A.~G. {Zhidkov}, A.~{Sasaki}, T.~{Tajima}, Direct spectroscopic observation of
  multiple-charged-ion acceleration by an intense femtosecond-pulse laser, {\it
  Physical Review E} {\bf 60}~(3) (1999), 3273--3278.

\bibitem{WangYang2007}
H.~{Wang}, X.~{Yang}, X.~{Li}, {Ground-State Energy Shifts of H-Like Ti Under
  Dense and Hot Plasma Conditions}, {\it Plasma Science and Technology} {\bf 9}
  (2007), 128--132.

\bibitem{Ashmore2011}
L.~{Ashmore}, Intrinsic plasma redshifts now reproduced in the laboratory---A
  discussion in terms of new tired light. {\it viXra:Astrophysics:1105.0010} .

\bibitem{Chen2009}
C.~S. {Chen}, X.~L. {Zhou}, B.~Y. {Man}, Y.~Q. {Zhang}, J.~{Guo},
  {Investigation of the mechanism of spectral emission and redshifts of atomic
  line in laser-induced plasmas}, {\it Optik} {\bf 120} (2009), 473--478.

\bibitem{Vogeley2012}
D.~C. {Pan}, M.~S. {Vogeley}, F.~{Hoyle}, Y.-Y. {Choi}, C.~{Park}, {Cosmic
  voids in Sloan Digital Sky Survey Data Release 7}, {\it \mnras} {\bf 421}
  (2012), 926--934.

\bibitem{Scheffler1987}
H.~{Scheffler}, H.~{Elsaesser}, {\it {Physics of the galaxy and interstellar
  matter}}, Springer-Verlag, Berlin, 1987.

\bibitem{Padmanabhan_III_2002}
P.~{Padmanabhan}, {\it {Theoretical astrophysics. Vol. III: Galaxies and
  Cosmology}}, {Cambridge University Press}, 2002.

\bibitem{Balakrishnan1991handbook}
N.~{Balakrishnan}, {\it Handbook of the Logistic Distribution}, Taylor \&
  Francis, New York, 1991.

\bibitem{univariate2}
N.~L. {Johnson}, S.~{Kotz}, N.~{Balakrishnan}, {\it Continuous univariate
  distributions}. Vol. 2. 2nd ed., {Wiley}, New York, 1995.

\bibitem{evans}
M.~{Evans}, N.~{Hastings}, B.~{Peacock}, {\it Statistical Distributions}, 3rd.
  ed., Wiley, New York, 2000.

\bibitem{Abramowitz1965}
M.~{Abramowitz}, I.~A. {Stegun}, {\it {Handbook of Mathematical Functions with
  Formulas, Graphs, and Mathematical Tables}}, Dover, New York, 1965.

\bibitem{Zaninetti2012e}
L.~{Zaninetti}, {New Analytical Results for Poissonian and non-Poissonian
  Statistics of Cosmic Voids}, {\it Revista Mexicana de Astronomia y
  Astrofisica} {\bf 48} (2012), 209--222.

\bibitem{Lane1870}
H.~J. {Lane}, On the theoretical temperature of the sun, under the hypothesis
  of a gaseous mass maintaining its volume by its internal heat, and depending
  on the laws of gases as known to terrestrial experiment, {\it American
  Journal of Science} ~(148) (1870), 57--74.

\bibitem{Emden1907}
R.~{Emden}, {\it Gaskugeln: Anwendungen der mechanischen W\"{a}rmetheorie auf
  kosmologische und meteorologische Probleme}, Teubner, Berlin, 1907.

\bibitem{Chandrasekhar_1967}
S.~{Chandrasekhar}, {\it {An Introduction to the Study of Stellar Structure}},
  New York, 1967.

\bibitem{Binney1987}
J.~{Binney}, S.~{Tremaine}, {\it {Galactic Dynamics}}, Princeton University
  Press, 1987.

\bibitem{Zwillinger1989}
D.~{Zwillinger}, {\it {Handbook of Differential Equations}}, Academic Press,
  New York, 1989.

\end{thebibliography}

\end{document}